\documentstyle[prb,aps,amsfonts,amssymb,floats,epsfig]{revtex}
\begin{document}
\draft 
\twocolumn[\hsize\textwidth\columnwidth\hsize\csname @twocolumnfalse\endcsname
\title{Optical spectroscopic study of the interplay 
of spin and charge in $\alpha^{\prime}$-NaV$_2$O$_5$}
\author{A. Damascelli\cite{byline}, C. Presura, and D. van der Marel}
\address{Solid State Physics Laboratory, University
 of Groningen, Nijenborgh 4, 9747 AG Groningen, The Netherlands}
\author{J. Jegoudez, and A. Revcolevschi} 
\address{Laboratoire de Chimie des Solides, Universit$\acute{e}$ de Paris-sud, 
B$\hat{a}$timent 414, F-91405 Orsay, France}
\date{May 10, 1999}
\maketitle
\begin{abstract}   
We investigate the temperature dependent optical properties of 
$\alpha^{\prime}$-NaV$_2$O$_5$, in the energy range 4\,meV-4\,eV. The symmetry of the 
system is discussed on the basis of infrared phonon spectra. 
By analyzing the optically  allowed phonons at temperatures below and above the phase 
transition, we conclude that a second-order change to a larger unit cell takes place below 34 K, 
with a fluctuation regime  extending over a broad temperature range.
In the high temperature undistorted phase, we find good agreement with the recently proposed 
centrosymmetric space group {\em Pmmn}. 
On the other hand, the detailed analysis of the electronic excitations detected in the 
optical conductivity, provides direct evidence for a charge disproportionated electronic 
ground-state, at least on a locale scale: A consistent interpretation of both structural and 
optical conductivity data requires an asymmetrical charge distribution on each rung, 
without any long range order. We show that, because of the locally broken symmetry, spin-flip 
excitations carry a finite electric dipole moment, which is responsible for the detection of 
direct two-magnon optical absorption processes for $E\!\parallel\!a$.
The charged-magnon model, developed to interpret the optical conductivity of 
$\alpha^{\prime}$-NaV$_2$O$_5$, is described in detail, and its relevance to other  
strongly correlated electron systems, where the interplay of spin and charge plays 
a crucial role in determining the low energy electrodynamics, is discussed. 
\end{abstract}
%
\vskip2pc]
\narrowtext

\section{Introduction}

After many years of intensive experimental and theoretical work on CuGeO$_3$ another inorganic 
compound, $\alpha^{\prime}$-NaV$_2$O$_5$, has been attracting the attention of the scientific 
community working on low-dimensional spin systems, in general, and on the spin-Peierls (SP) 
phenomenon, in particular. In fact, in 1996 the isotropic activated  behavior of the magnetic 
susceptibility has been observed at low temperatures on $\alpha^{\prime}$-NaV$_2$O$_5$.\cite{isobe} 
Moreover, superlattice reflections, 
with a lattice modulation vector $k\!=\!(\pi\!/\!a,\pi\!/b,\pi\!/2c)$, were found in an x-ray 
scattering experiment,\cite{fujii} and a spin gap $\Delta$=9.8 meV was observed, at the reciprocal 
lattice point $(2\pi\!/\!a,\pi\!/b,0)$, with inelastic neutron scattering.\cite{fujii} The SP 
picture was then proposed to explain the low temperature properties of this compound, with a SP 
transition temperatures $T_{\rm{SP}}$=34 K.\cite{isobe} In this context, 
$\alpha^{\prime}$-NaV$_2$O$_5$ seemed to be a particularly interesting  material, as far as the 
full understanding of the SP phenomenon is concerned, because on the basis of the structural 
analysis reported by Carpy and Galy in 1975, suggesting the noncentrosymmetric space group 
$P2_1mn$,\cite{carpy} it was assumed to be described by the one-dimensional (1D) spin-1/2 
Heisenberg (HB) model better than CuGeO$_3$. In fact, for the latter compound it has been 
established both experimentally\cite{nishi} and theoretically\cite{uhrig} that the 2D character 
of the system cannot be neglected.

On the other hand, on the basis of experimental results later obtained, the 
interpretation of the phase transition  is still controversial. 
First of all, the reduction of $T_{\rm{SP}}$ upon increasing the intensity of an externally 
applied magnetic field, expected for a true SP system, has not been 
observed.\cite{buchner,schnelle} For instance, B\"uchner {\em et al.},\cite{buchner} performing 
low temperature magnetization and specific heat measurements in magnetic field $H\!\leq\!14$\,T, 
obtained $\Delta T_{\rm{SP}}(H)\!=\!T_{\rm{SP}}(0)\!-\!T_{\rm{SP}}(14\, \rm{T})\!\approx\!0.15$ K, 
i.e., almost a factor of 7 smaller that what expected on the basis of Cross and Fisher 
theory.\cite{crossfi,cross} Furthermore, they claimed that the entropy reduction experimentally 
observed across the phase transition is considerably larger than the theoretical expectation 
for a 1D antiferromagnetic (AF) HB chain with only spin degrees of freedom and $J\!=\!48$ meV 
(i.e., value obtained for $\alpha^{\prime}$-NaV$_2$O$_5$ from magnetic susceptibility 
measurements\cite{isobe}). In order to explain these findings, 
degrees of freedom additional to those of the spin system should be taken into account, contrary 
to the interpretation of a magnetically driven phase transition.\cite{buchner} 

This already quite puzzling picture became even more complicated when the crystal structure and 
the symmetry of the electronic configuration of $\alpha^{\prime}$-NaV$_2$O$_5$, in the high 
temperature phase, were investigated in detail. In fact, on the basis of new x-ray diffraction 
measurements\cite{meetsma,schnering,smolinski} it was found that the 
symmetry of this compound at room temperature is better described by the centrosymmetric space 
group $Pmmn$ than by the originally proposed $P2_1mn$.\cite{carpy} On the other hand, the 
intensities and polarization dependence of the electronic excitations detected in optical 
conductivity spectra gave a direct evidence for a broken-parity electronic ground 
state,\cite{anvprl,anvphysica}  in apparently better agreement with the noncentrosymmetric space 
group $P2_1mn$. The solution of this controversy and 
the assessment of the symmetry issue are of fundamental importance for the understanding of the 
electronic and magnetic properties of the system and, ultimately, of the phase transition. 

In this paper we present a detail investigation of the interplay of spin and charge in 
$\alpha^{\prime}$-NaV$_2$O$_5$, on the basis of optical reflectivity and conductivity data in 
the energy range 4\,meV-4\,eV. We will start from the discussion of the newly proposed high 
temperature crystal structure,\cite{meetsma,schnering,smolinski} and its implications on the 
physics of this compound (section\ \ref{nvrtcs}). A group theoretical analysis 
for the two different 
space groups will be reported and later compared to the experimentally observed far-infrared phonon 
spectra. In section\ \ref{chmgmod}, we will present a theoretical model which we will use to 
interpret qualitatively and quantitatively the optical conductivity spectra of 
$\alpha^{\prime}$-NaV$_2$O$_5$. In section\ \ref{nvoptsr},  we will analyze the data obtained 
with optical spectroscopy at different temperatures above and below 
$T_{\rm{c}}\!=\!34$ K (throughout 
the paper, we will refer to the 
transition temperature as $T_{\rm{c}}$ because the interpretation of the 
phase transition is still controversial). In particular, we will concentrate on:

(\/{\em i}\,) Analysis of the temperature dependent phonon spectra, in relation to the x-ray 
diffraction data, in order to: First, learn more about the symmetry of the material in both high 
and low temperature phase. Second, detect signatures of the lattice distortion and study the 
character of the phase transition.
 
(\/{\em ii}\,) Detailed study of the very peculiar electronic and magnetic excitation spectra. In 
particular, we will show that we could detect a low frequency continuum of excitations which, 
on the basis of the model developed in section\ \ref{chmgmod}, we ascribe to `charged bi-magnons', 
i.e., direct two-magnon optical absorption processes. 

Finally in section\ \ref{nvdisc}, we will discuss the relevance of our findings to the 
understanding of the nature of the phase transition in $\alpha^{\prime}$-NaV$_2$O$_5$. 
In particular we will try to assess whether the picture of a charge ordering transition, 
accompanied by the opening of a magnetic gap, is a valid alternative to the originally proposed 
SP description. In fact, this 
alternative interpretation has been recently put forward,\cite{ohama,seo,thalmeier,mostovoy,riera} 
on the basis of temperature dependent nuclear magnetic resonance (NMR) data.\cite{ohama}

\section {The Symmetry Problem in $\alpha^{\prime}$-NaV$_2$O$_5$}
\label{nvrtcs}

\begin{figure}[t]
\centerline{\epsfig{figure=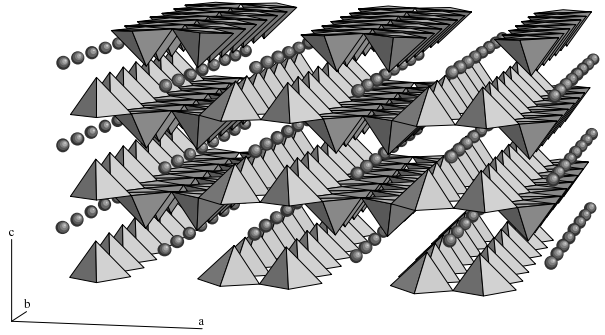,width=8cm,clip=}}
\vspace{0.3cm}
\caption{High temperature crystal structure of $\alpha^{\prime}$-NaV$_2$O$_5$: $\!$Pyramids 
around V ions and rows of Na ions are shown.}
\label{nv3d}
\end{figure}

\begin{figure}[b]
\centerline{\epsfig{figure=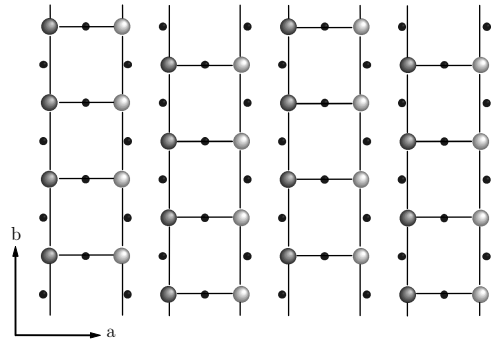,width=6.4cm,clip=}}
\vspace{0.3cm}
\caption{Two-leg ladder structure formed by O and V ions in the $a$-$b$ plane of 
$\alpha^{\prime}$-NaV$_2$O$_5$. Black dots are O ions; dark and light gray balls are 
V$^{4+}$ and V$^{5+}$ ions, respectively, arranged 
in 1D chains in the space group $P2_1mn$.}
\label{nv2d}
\end{figure}

The interpretation of the phase transition at 34 K as a SP transition\cite{isobe} is based on the 
noncentrosymmetric space group $P2_1mn$ originally proposed by Carpy and Galy for the high 
temperature phase of $\alpha^{\prime}$-NaV$_2$O$_5$.\cite{carpy}  Following their crystallographic 
analysis, the structure can be constructed from double-rows (parallel to the {\em b} axis) of edge 
sharing pyramids, one facing up and the other down, with respect to the {\em a-b} plane 
(see Fig.\ \ref{nv3d}). These double rows are connected by sharing corners, yielding a planar 
material. The planes are stacked along the {\em c} axis, with the Na in the channels of the 
pyramids. In the {\em a-b} plane (see Fig.\ \ref{nv2d}), we can then identify linear chains of 
alternating V and O ions, oriented along the {\em b} axis. These chains are grouped into sets of 
two, forming a ladder, with the rungs oriented along the {\em a} axis. The rungs are formed by two 
V ions, one on each leg of the ladder, bridged by an O ion. The V-O distances along the rungs 
are shorter than along the legs, implying a stronger bonding along the rung. In the {\em a-b} 
plane the ladders are shifted half a period along the {\em b} axis relative to their neighbors.
The noncentrosymmetric space group $P2_1mn$, allows for two inequivalent V positions in the 
asymmetric unit. These sites were interpreted as different valence states, V$^{4+}$ and V$^{5+}$, 
represented by dark and light gray balls, respectively, in Fig.\ \ref{nv2d}. In this structure 
it is possible to identify well-distinct 1D magnetic V$^{4+}$ (S=1/2) and non-magnetic V$^{5+}$ 
(S=0) chains running along the {\em b} axis of the crystal, and alternating each other along the 
{\em a} axis. This configuration would  be responsible for the 1D character 
(Bonner-Fisher-type\cite{bonner}) of the high temperature susceptibility\cite{mila} and for the 
SP transition, possibly involving dimerization within the V$^{4+}$ chains.\cite{isobe} Moreover, 
due to the details of the crystal structure, only one of the three $t_{2g}$ orbitals of V$^{4+}$, 
i.e., the $d_{xy}$ orbital, is occupied.\cite{smolinski} As a result, the insulating character of 
$\alpha^{\prime}$-NaV$_2$O$_5$, which has been observed in DC resistivity 
measurements,\cite{isobe1,hemberger} would be an obvious consequence of  having a 1D 1/2-filled 
Mott-Hubbard system.  
%

However, recent redeterminations of the crystal structure\cite{meetsma,schnering,smolinski} showed 
that the symmetry of $\alpha^{\prime}$-NaV$_2$O$_5$ above $T_{\rm{c}}$ is better 
described by the centrosymmetric space group {\em Pmmn}: The suggested structure is very similar to 
the one proposed by  Carpy and Galy\cite{carpy} (a small difference is the somewhat more 
symmetric eight-fold  coordination of Na\cite{meetsma}). 
On the other hand, one major difference was found, namely, the presence of a center of inversion 
in the unit cell. The evidence is the very low $R(F)$-value 
of 0.015, and the fact that no lower value of $R(F)$ can be obtained when the x-ray diffraction 
data are refined omitting the inversion center.\cite{meetsma} In particular, this result 
suggests the existence of only one kind of V per unit cell, with an average valence for the V 
ions of +4.5. This finding has important implications for the understanding of the electronic and 
magnetic properties of $\alpha^{\prime}$-NaV$_2$O$_5$. In fact, we have now to think in the 
framework of a 1/4-filled two-leg ladder system: It is still possible to recover an insulating 
ground state assuming that the {\em d} electron, supplied by the two V ions forming a rung of 
the ladder, is not attached to a particular V site but is shared in a V-O-V molecular bonding 
orbital along the rung.\cite{smolinski} In this way, we are effectively back to a 1D spin system 
(a two-leg ladder with one spin per rung in a symmetrical position). However, in case of a SP 
phase transition, singlets would have to be formed by electrons laying in molecular-like orbitals 
and a rather complicated distortion pattern would have to take place: Not simply the dimerization 
within a real 1D chain (as, e.g., in CuGeO$_3$), but a deformation of the plaquettes, formed by 
four V ions, within a ladder.

Regarding the controversy about the x-ray structure determination for 
$\alpha^{\prime}$-NaV$_2$O$_5$ above $T_{\rm{c}}$, it is important to stress that, 
even though it is now well established that the `most probable' space group is the centrosymmetric 
{\em Pmmn} (on the basis of the statistics of the refinement), 
deviations from centrosymmetry up to 0.029 \AA\  are unlikely but cannot be excluded.\cite{meetsma}
Moreover, one has to note that x-ray diffraction measurements are sensitive to the charge 
distribution of core electrons and not of valence electrons. Therefore, if a local breaking of 
symmetry in the distribution of the V {\em d} electrons were present on a local scale without the 
long range order required by the noncentrosymmetric space group $P2_1mn$ (as suggested by the 
optical conductivity data we will present in section\ \ref{nvoc}), it would not be detectable in 
an x-ray diffraction experiment: X-rays would just see the 'space-average' given by the 
centrosymmetric space group {\em Pmmn}.

\subsection{Group Theoretical Analysis}
\label{nvgta}

An alternative way to assess the symmetry issue is to investigate the Raman  and infrared phonon 
spectra, comparing the number of experimentally observed modes to the number expected for the 
two different space groups on the basis of group-theory. Obviously, 
a large number of modes has to be expected because each unit cell contains two formula units,  
corresponding to 48 degrees of freedom, which will be reflected in 48 phonon branches.

In the space group {\em Pmmn}  the site groups in the unit cell are: $C_{2v}^{z}$ 
for the 2 Na and the 2 O(1) atoms, and $C_{s}^{xz}$ for the 4 V, the 4 O(2), and the 4 O(3). 
Following the nuclear site group analysis method extended to crystals,\cite{rousseau} 
the contribution of each occupied site to the total irreducible representation of the crystal is:
\begin{eqnarray}
  &&\Gamma_{\rm{Na+O(1)}}\!=\!2[A_{g}\!+\!B_{1u}\!+\!B_{2g}\!
                        +\!B_{2u}\!+\!B_{3g}\!+\!B_{3u}]\nonumber \, , \\
  &&\Gamma_{\rm{V+O(2)+O(3)}}\!=\!3[2A_{g}\!+\!A_{u}\!+\!B_{1g}\!+\!2B_{1u} \nonumber \\
  &&\hspace{4cm}+2B_{2g}\!+\!B_{2u}\!+\!B_{3g}\!+\!2B_{3u}]\nonumber \, .  
\end{eqnarray}
Subtracting the silent modes (3$A_{u}$) and the acoustic modes 
($B_{1u}\!+\!B_{2u}\!+\!B_{3u}$), the irreducible representation of the optical vibrations, 
for the centrosymmetric space group {\em Pmmn}, is:
\begin{eqnarray}
 \Gamma\!=\!&&\,8A_{g}(aa,bb,cc)\!+\!3B_{1g}(ab)\!+\!8B_{2g}(ac)\!+\!5B_{3g}(bc) \nonumber \\
         &&+7B_{1u}(E\|c)\!+\!4B_{2u}(E\|b)\!+\!7B_{3u}(E\|a)  \, ,
\label{nvrepme}
\end{eqnarray}
corresponding to 24 Raman ($A_{g}$,$B_{1g}$,$B_{2g}$,$B_{3g}$) and 18 infrared
($B_{1u}$,$B_{2u}$,$B_{3u}$) active modes.

In the analysis for the noncentrosymmetric space group $P2_1mn$, we have to consider 
only the  site group $C_{s}^{yz}$ for all the atoms in the unit cell: 2\,Na, 2\,V(1), 2\,V(2), 
2\,O(1), 2\,O(2), 2\,O(3), 2\,O(4), and 2\,O(5). Therefore, the total irreducible representation 
is:
\begin{eqnarray}
  8\times\Gamma_{C_{s}^{yz}}\!=\!8\!\times\![2A_{1}\!+\!A_{2}\!+\!B_{1}\!+\!2B_{2}] \nonumber \, .
\end{eqnarray}
Once again, subtracting the acoustic modes  $A_{1}\!+\!B_{1}\!+\!B_{2}$ (there is no silent mode 
in this particular case), the irreducible representation of the optical vibrations, 
for the noncentrosymmetric space group $P2_1mn$, is: 
\begin{eqnarray}
    \Gamma^{\prime}\!=\!&&15A_{1}(aa,bb,cc;E\|a)\!+\!8A_{2}(bc) \nonumber \\
    &&+7B_{1}(ab;E\|b)\!+\!15B_{2}(ac;E\|c) \,\, ,
\label{nvrepcg}
\end{eqnarray}
corresponding to 45 Raman ($A_{1}$,$A_{2}$,$B_{1}$,$B_{2}$) and 37 infrared 
($A_{1}$,$B_{1}$,$B_{2}$) active modes. 

As a final result, we see that the number of optical vibrations, 
expected on the basis of group theory,  is very different for the two space groups. Moreover, 
whereas in the case of {\em Pmmn} the phonons are exclusively Raman or infrared active, for 
$P2_1mn$, because of the lack of inversion symmetry, there is no more distinction between 
{\em gerade} and {\em ungerade} and certain modes are, in principle (group theory 
does not say anything about intensities), detectable with both techniques.

\section{Charged-Magnon Model}
\label{chmgmod}

In the context of the extensive work done, in the last few years, on 2D S=1/2 quantum AF and 
their 1D analogs, it has been shown both experimentally and 
theoretically\cite{perkins,lorenzana1,lorenzana2,suzuura,lorenzana3,kastner} 
that also optical spectroscopy, besides elective techniques like Raman and 
neutron scattering, can be a very useful probe in studying the spin dynamics in this systems. 
For example, in the 2D-system La$_2$CuO$_4$ 
(1D-system Sr$_2$CuO$_3$) where two-magnon (two-spinon) excitations are in principle not 
optically active because of the presence of a center of inversion that inhibits the formation 
of any finite dipole moment, phonon-assisted magnetic excitations were detected in the 
mid-infrared region.\cite{perkins,suzuura} These magneto-elastic absorption processes are 
optically allowed because the phonon there involved is effectively lowering the symmetry of 
the system.

Let us now concentrate on a system where a breaking of symmetry is present 
because of charge ordering. We show that, in this case, magnetic excitations are expected to 
be directly optically active and detectable in an optical experiment as {\em charged bi-magnons}: 
Double spin-flip excitations carrying a finite dipole moment. We will do that investigating the 
spin-photon interaction in a single two-leg ladder with a charge disproportionated ground state. 
On the basis of this  model we will later interpret  the optical 
conductivity spectra of $\alpha^{\prime}$-NaV$_2$O$_5$. 

\subsection{Model Hamiltonian}
\label{modh}

\begin{figure}[b]
\centerline{\epsfig{figure=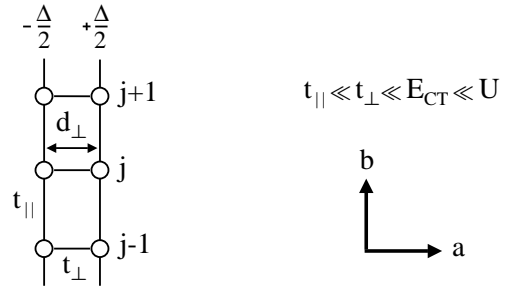,width=6.5cm,clip=}}
\vspace{0.3cm}
\caption{Sketch of the single two-leg ladder. Circles represent the ionic 
sites having on-site energies -$\Delta$/2 and +$\Delta$/2 on the left and on the right leg of 
the ladder, respectively. $t_{\perp}$ and 
$t_{\parallel}$ are the hopping parameters for the rungs and the legs, respectively, 
$d_{\perp}$ is the length of a rung, and $j$ is the rung index.  Also indicated is the relation,  
assumed throughout the discussion, between the different energy scales of the model.} 
\label{stleg} 
\end{figure}

In this section we discuss the single two-leg ladder depicted in Fig.\ \ref{stleg}, 
where $t_{\perp}$ and $t_{\parallel}$ are the hopping parameters for the rungs and the legs, 
respectively, $d_{\perp}$ is the length of a rung, and -$\Delta$/2 and +$\Delta$/2 are the 
on-site energies for the left and the right leg of the ladder, respectively. We work at quarter 
filling, i.e., one electron per rung. The total Hamiltonian $H_T$ of the system is:
\begin{eqnarray}
      H_T &&= H_0 + H_{\bot} + H_{\parallel}   \\ 
      \nonumber \\
      H_{0\,} &&= U \sum_{j}\left\{n_{jR\uparrow}n_{jR\downarrow} 
               + n_{jL\uparrow}n_{jL\downarrow}\right\} + \Delta \sum_{j} n_{jC} \label{h0}  \\
H_ {\bot\!} &&= t_{\bot} \sum_{j,\sigma}\left\{L_{j,\sigma}^{\dagger}R_{j,\sigma} 
                + H.c.\right\} \label{hperp} \\
H_ {\parallel\,} &&= t_{\parallel} \sum_{j,\sigma}\left\{R^{\dagger}_{j,\sigma}R_{j+1,\sigma} 
                      + L^{\dagger}_{j,\sigma}L_{j+1,\sigma} + H.c. \right\} \label{hpara} \,  , 
\end{eqnarray}  
where $L_{j,\sigma}^{\dagger}$ ($R_{j,\sigma}^{\dagger}$) creates an electron 
with spin $\sigma$ on the left-hand (right-hand) site of the {\em j}\,th rung, $U$ is the on-site 
Hubbard repulsion, $n_{jR\uparrow}n_{jR\downarrow}$ ($n_{jL\uparrow}n_{jL\downarrow}$) is 
counting the double occupancies on the right-hand (left-hand) site,  
$n_{jC}\!=\!(n_{jR}-n_{jL})/2$ is the 
charge displacement operator, and $\Delta$ is the potential energy difference between the two 
sites. Considering only $H_0$ at quarter filling one immediately realizes that  $\Delta$ is not 
only breaking the left-right symmetry of the system (for the symmetric ladder $\Delta\!=\!0$), 
but it is also 
introducing a long range charge ordering with one electron per site on the left leg of the ladder. 
Working in the limit $t_{\bot}\!\gg\!t_{\parallel}$ we can consider each rung as an independent 
polar molecule (with one electron) described by the Hamiltonian $H_{j0}\!+\!H_{j\bot}$. The two 
solutions are lob-sided bonding and antibonding 
wave functions $|\tilde{L}\rangle=\!u|L\rangle\!+v|R\rangle$, and $|\tilde{R}\rangle\!=
u|R\rangle\!-v|L\rangle$, with:
\begin{equation}
 u = {\frac{1} {\sqrt{2}}} \sqrt{1+ \frac{\Delta} {E_{CT}} }\,  , \hspace{1cm}
 v = {\frac{1} {\sqrt{2}}} \sqrt{1- \frac{\Delta} {E_{CT}} }\,  , 
\label{chmcoeff}
\end{equation} 
where $E_{CT}\!=\!\sqrt{\Delta^2+4t_{\bot}^2}$ is the splitting between these two eigenstates. 
The excitation of the electron from the bonding to the antibonding state is an optically active 
transition with a  degree of charge transfer (CT) from the left to the right site which is the 
larger the bigger is $\Delta$. It is possible to calculate the  integral of the real part of the 
 optical conductivity $\sigma_1(\omega)$ for this excitation, quantity that can be very useful 
 in the analysis of optical spectra. We start from the general equation:\cite{ziman}
\begin{equation}
\int_{0}^{\infty} \sigma_{1}(\omega) d\omega =  \frac {\pi q_e^2} {\hbar^{2} V} 
  \sum_{n\neq g} \left( E_n-E_g \right) \mid\!\!\langle n|\sum_{i} {\bf x}_i 
  |g\rangle\!\!\mid^{2} \,  , 
\label{siggen}
\end{equation}
where {\em V} is the volume, $q_e$ is the electron charge and {\em i} is the site index.  
As for a single rung $\langle \tilde{R}|{\bf x}_i|\tilde{L}\rangle\!=\!-d_{\bot}t_{\bot}/E_{CT}$, we 
obtain the following expression, which is exact for one electron on two coupled tight-binding 
orbitals:
 \begin{equation}
 \int_{CT} \sigma_{1}(\omega) d\omega = 
 \pi q_e^2 N d_{\bot}^2 t_{\bot}^2 \hbar^{-2} E_{CT}^{-1}\,  ,
 \label{sigCT}
 \end{equation}
where {\em N} is the volume density of the rungs. Comparing Eq.\ \ref{sigCT} and the expression 
for 
$E_{CT}$ with the area and the energy position of the CT peak observed in the experiment, we can 
extract both $t_{\bot}$ and $\Delta$.

In the ground state of the Hamiltonian $H_{0}\!+\!H_{\bot}$, each electron resides in a
$|L\rangle$ orbital, with some admixture of $|R\rangle$. Let us now introduce the coupling of the 
rungs along the legs, considering $H_{\parallel}$ and a small fragment of the ladder with only 
two rungs. We have now to take into account the spin degrees of freedom: If the two spins are 
parallel, the inclusion of $H_{\parallel}$ has no effect, due to the Pauli-principle. If they 
form a $S=0$ state, the ground state can gain some kinetic energy along the legs by
mixing in states where two electrons reside on the same rung. We start from the singlet 
ground state for the two rungs:
\begin{equation}
 |\tilde{L}_1 \tilde{L}_2\rangle= \frac {1} {\sqrt{2}} \left\{ 
|\tilde{L}_{1\uparrow} 
 \tilde{L}_{2\downarrow} \rangle - 
 |\tilde{L}_{1\downarrow} \tilde{L}_{2\uparrow} \rangle \right\}\,  ,
 \end{equation}
with eigenvalue $E_{L_1L_2}\!=\!-E_{CT}$, and we study the effect of $H_{\parallel}$ using a 
perturbation expansion in $t_{\parallel}/E_{CT}$, to the second order. However, 
$H_{\parallel}|\tilde{L}_1 \tilde{L}_2\rangle$ does not coincide with any of the singlet 
eigenstates of the system consisting of two electrons on one rung with the Hamiltonian 
$H_{j0}\!+\!H_{j\bot}$. In order to obtain the exchange coupling constant 
$J_{\parallel}$ between two spins on neighboring rungs, we have first to calculate the set of 
spin-singlet eigenstates for the two electrons on one rung, and then to evaluate the fractional 
parentages of $H_{\parallel}|\tilde{L}_1 \tilde{L}_2\rangle$ with respect to that set.
The Hilbert subspace of two-electron spin-singlet states is spanned by the vectors:
\begin{eqnarray}
 |\,L_jL_j\rangle &&=  L_{j,\uparrow}^{\dagger} L_{j,\downarrow}^{\dagger}|0\rangle 
\\   
 |L_jR_j\rangle &&= \frac {1} {\sqrt{2}} \left\{ L_{j,\uparrow}^{\dagger} 
 R_{j,\downarrow}^{\dagger}- L_{j,\downarrow}^{\dagger} R_{j,\uparrow}^{\dagger} \right\}|0\rangle 
\\      
 |R_jR_j\rangle &&=  R_{j,\uparrow}^{\dagger} R_{j,\downarrow}^{\dagger}|0\rangle \, .
\end{eqnarray}
On this basis the Hamiltonian of the {\em j}\,th rung is:
\begin{equation}
 H_{j0}+H_{j\perp} = \left[\matrix{ 
       U-\Delta & \sqrt{2}t_{\perp} & 0   \cr  
       \sqrt{2}t_{\perp} & 0 & \sqrt{2}t_{\perp}  \cr 
       0 & \sqrt{2}t_{\perp} & U+\Delta}\right] \, .
\end{equation}
A significant simplification of the problem occurs in the large $U$ limit which, moreover, 
elucidates the physics of exchange processes between a pair of dimers in an elegant and simple way. 
We therefore take the limit $U\!\rightarrow\!\infty$. The solutions for $\Delta\!=\!0$ 
are:\cite{mjr79} The (anti)-symmetric linear combinations of $|L_jL_j\rangle$ and 
$|R_jR_j\rangle$ with energy $U$, and $|L_jR_j\rangle$ with energy 
$E_{LR}\!=\!-4t_{\perp}^2/U\!\rightarrow\!0$. The triplet states have energy 0. 
For a general value of $\Delta$, still obeying $\Delta\!\ll\!U$, the only {\em relevant} 
eigenstate for the calculation of the exchange processes between neighboring dimers is 
$|L_jR_j\rangle$, with energy $E_{LR}\!=\!0$. The other two states with energy of order 
$U$ are projected out in this limit. 

We can now go back to the problem of two rungs with one electron per rung, and consider the 
hopping along the legs treating $H_{\parallel}$ as a small perturbation  with respect to 
$H_{0}\!+\!H_{\bot}$. We proceed by calculating the corrections to 
$|\tilde{L}_1 \tilde{L}_2\rangle$ by allowing a finite hopping parameter $t_{\parallel}$ 
between the rungs. Using lowest order perturbation theory, the ground state energy of a 
spin-singlet is:
\begin{equation}
 E_{g,S=0} = -E_{CT} - \langle\tilde{L}_1 \tilde{L}_2| H_{\parallel}    
  \frac{1}{ H_{0}\!+\!H_{\perp}\!+\!E_{CT}}  H_{\parallel} |\tilde{L}_1 \tilde{L}_2\rangle  \,.
\end{equation}                                                                              
We proceed by calculating the coefficients of fractional parentage by projecting 
$H_{\parallel}|\tilde{L}_1 \tilde{L}_2\rangle$ on the two-electron single eigenstates of 
a single rung:
\begin{eqnarray}
 H_{\parallel}|\tilde{L}_1 \tilde{L}_2\rangle = \sqrt{2}\,t_{\parallel}\sum_{j=1,2} 
               && \{ u^2 |L_jL_j\rangle\!+\!v^2 |R_jR_j\rangle \nonumber\\
  &&+\sqrt{2}\,u v\, |L_jR_j\rangle \} \,.
\end{eqnarray}
Calculating the second order correction to $E_{L_1L_2}$ for the $S$=0 state, in the limit where 
$U\!\rightarrow\!\infty$, and realizing that the correction for the $S$=1 spin configuration is 
zero, we obtain the exchange coupling constant $J_{\parallel}$:
\begin{equation}
 J_{\parallel} = \frac{8\,t_{\parallel}^2 \, t_{\perp}^2} 
 {\left[\Delta^2+4\,t_{\perp}^2 \right]^{3/2}} \,.
\label{jpara}
\end{equation}

\subsection{Interaction Hamiltonians and Effective Charges}

\begin{figure}[t]
\centerline{\epsfig{figure=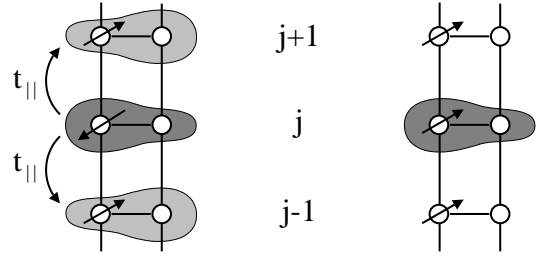,width=7cm,clip=}}
\vspace{0.3cm}
\caption{Pictorial description of the electric dipole moment associated with a single 
spin-flip in the asymmetrical two-leg ladder. If spins are antiparallel (left-hand side), 
the charge of the  $j$\,th electron is asymmetrically distributed not only over the $j$\,th rung 
but also on the nn rungs (with opposite asymmetry), because of the virtual hopping. If we flip 
its spin (right-hand side), the $j$\,th electron is confined to the  $j$\,th rung, resulting in a 
reduced charge density on the right leg of the ladder. Therefore, there is a net dipole 
displacement (along the rung direction) of the antiparallel spin configuration compared to the 
parallel one.} 
\label{sflip} 
\end{figure}

Let us now consider three rungs of the ladder (see Fig.\ \ref{sflip}) to understand the role of spin-flip 
excitations in this system. Because of the energy gain $J_{\parallel}$, in the ground state we 
have antiparallel alignment of the spins (Fig.\ \ref{sflip}, left-hand side). 
The {\em j}\,th 
electron is described by the lob-sided bonding wave function  $|\tilde{L}_j\rangle$ and, 
because  of the virtual hopping to the nearest-neighbor (nn) rungs, by states where two electrons 
reside on the same rung. Working in the limit $U\!\rightarrow\!\infty$ these states 
have one electron in the $|L\rangle$ and one in the $|R\rangle$ state on the same rung. 
Therefore, the charge of the {\em j}\,th electron is asymmetrically distributed not only over the 
{\em j}\,th rung (dark grey area in Fig.\ \ref{sflip}, left) but also on the nn rungs 
(light grey area in Fig.\ \ref{sflip}, left). However, in the latter case most of the charge 
density is localized on the right leg of the ladder. 
If we  flip the spin of the  {\em j}\,th electron (Fig.\ \ref{sflip}, right-hand side)  
no virtual hopping is possible any more to the nn rungs because of the Pauli 
principle. The electron charge distribution is now determined only by  the lob-sided  wave 
function  $|\tilde{L}_j\rangle$ (dark grey area in Fig.\ \ref{sflip}, right). As a result, 
there is a net dipole displacement of the antiparallel spin configuration compared to the parallel
one: The spin-flip excitations carry a finite dipole moment parallel to the rung direction 
({\em a} axis). This dipole moment can couple to the electric field of the incident radiation in 
an optical experiment, and result in what we refer to as {\em charged bi-magnon} excitations. 
One may remark, at this point, that only double 
spin flips can be probed in an optical experiment because of spin conservation. However, what we 
just discuss for a single spin flip can be extended to the situation where two spin flips, on 
two different rungs, are considered.

The coupling to  light with  $\vec{E}\!\parallel\!\vec{a}$ (rung direction) 
can  now be included   using the dipole approximation.  The
only effect is to change the potential energy of the $|R\rangle$ states relative
to the $|L\rangle$ states. In other words, we have to replace $\Delta$ with
$\Delta+q_e d_{\bot} E_a$, where  
$E_a$ is the component of the electric field along the rung. The coupling to the CT transition 
on the rungs is given by the Hamiltonian:
\begin{equation}
  H_{CT}=q_e d_{\bot} E_a n_{C}=q_e d_{\bot} E_a \sum_{j}n_{jC} \, .
 \end{equation}
The Hamiltonian for the spin-photon coupling can be obtained taking the Taylor expansion of 
$J_{\parallel}$ with respect to $E_a$, and retaining the term which is linear in the electric 
field. Noticing that $dE_a\!=\!d\Delta/q_ed_{\bot}$, we have:
 \begin{equation}
  H_S=q_m d_{\bot} E_a h_{S}=q_m d_{\bot} E_a \sum_{j}\vec{S}_j\cdot\vec{S}_{j + 1} \, ,
 \end{equation}
where $q_m\!=\!\frac {1} {d_{\bot}} \frac {\partial J_{\parallel}} {\partial E_a}\!=\!
 q_e \frac {\partial J_{\parallel}} {\partial \Delta}$
is the effective charge involved in a double  spin-flip transition, and 
$h_{S}\!=\!\sum_{j}\vec{S}_j\!\cdot\!\vec{S}_{j + 1}$. 
In the limit $U\!\rightarrow\!\infty$, the effective charge reduces to:
\begin{equation}
 q_m\!=\!q_e \frac{3J_{\parallel}\Delta}{\Delta^2+4t_{\perp}^2}  \, ,
\label{efcgmg}
\end{equation}
where $J_{\parallel}$ is given by Eq.\ \ref{jpara}. One has to note that for a symmetrical 
ladder, where $\Delta\!=\!0$, the
effective charge $q_m$ vanishes, and the charged magnon effect disappears: The peculiar behavior
of the spin flips we just described is an exclusive consequence of the broken left-right symmetry 
of the ground state. On the other hand, the bonding-antibonding transition  would still be 
optically active. However, the bonding-antibonding energy splitting would be determined only 
by $t_{\bot}$, i.e., $E_{CT}^{\prime}\!=\!2t_{\bot}$. At this point we like to speculate on 
the role 
played by $\Delta$ in determining not only the optical activity of the double spin-flip 
excitations but also the insulating (metallic) nature of the system along the leg direction. 
For $\Delta\!=\!0$ and for strong on-site Coulomb repulsion $U$, the energy cost to transfer 
one electron from one rung to the neighboring one is $E_{CT}^{\prime}\!=\!2t_{\bot}$.  
Therefore, the 
system is insulating for $4t_{\parallel}\!<\!E_{CT}^{\prime}\!=\!2t_{\bot}$  and metallic in 
the opposite 
case. If we now start from a metallic situation 
$4t_{\parallel}\!\geq\! E_{CT}^{\prime}\!=\!2t_{\bot}$, 
switching 
on $\Delta$ we can realize the condition $4t_{\parallel}\!<\!E_{CT}\!=\!\sqrt{\Delta^2+4t_{\bot}^2}$ 
which would result in a metal-insulator transition and, at the same time, in optical activity 
for the charged bi-magnons.

\subsection{Spectral Weights}
\label{spwe}

In this section we will attempt to estimate what would be, in an optical experiment, the 
relative spectral weights for the CT and 
the charged bi-magnon excitations of the single two-leg ladder. In particular, we want to 
show that these quantities can be expressed in terms of properly defined correlation functions. 
Considering the integral of the optical conductivity, Eq.\ \ref{siggen}, and the interaction 
Hamiltonians $H_{CT}$ and $H_{S}$, we can write:
\begin{eqnarray}
\int_{CT} \sigma_{1}(\omega) d\omega &&=  \frac {\pi q_e^2 d_{\bot}^2} {\hbar^{2} V} 
  \sum_{n\neq g} \left( E_n-E_g \right) 
                \mid\!\!\langle n|n_C |g\rangle\!\!\mid^{2} \, , \label{sigCT1}\\
\int_{\;S\,\;} \sigma_{1}(\omega) d\omega &&=  \frac {\pi q_m^2 d_{\bot}^2} {\hbar^{2} V}
  \sum_{n\neq g} \left( E_n-E_g \right) \mid\!\!\langle n|h_S |g\rangle\!\!\mid^{2} \, . 
\label{sigS}
\end{eqnarray}

Let us first concentrate on the CT excitation which is a simpler problem because, as we 
discussed in section\ \ref{modh}, on each rung we are dealing with one electron on two 
coupled tight-binding orbitals $|\tilde{L}\rangle$ and $|\tilde{R}\rangle$. Therefore, only one 
energy term $(\!E_n\!-\!E_g)\!=\!E_{CT}$ has to be considered in the previous equation. We can then write:
\begin{eqnarray}
  &&\sum_{n\neq g} \left( E_n-E_g \right) \mid\!\!\langle n|n_C |g\rangle\!\!\mid^{2}= \nonumber\\
  &&=E_{CT} \left\{ \sum_n \mid\!\!\langle n|n_C |g\rangle\!\!\mid^{2} 
         - \mid\!\!\langle g|n_C |g\rangle\!\!\mid^{2} \right\}=  \nonumber\\
  &&=E_{CT} \left\{ \langle g|n_C^2 |g\rangle\ 
         - \mid\!\!\langle g|n_C |g\rangle\!\!\mid^{2} \right\} =  \nonumber\\ 
  &&=E_{CT} \left\{  \langle n_C^2 \rangle - \langle n_C \rangle^{2} \right\}  \,  , 
\label{sommona}
\end{eqnarray}
where we took into account that the expectation value of $n_C$ over 
the ground state is a real number. From Eq.\,\ref{sigCT1} and\,\ref{sommona}, we obtain:
\begin{equation}
\int_{CT} \sigma_{1}(\omega) d\omega =  \frac {\pi q_e^2 d_{\bot}^2} {4 \hbar^{2}V} 
  E_{CT} \, g_C(T) \,  , 
\label{sigCTgc}
\end{equation}
where we have introduced the correlation function: 
\begin{equation}
g_C(T)= 4 \mbox{\boldmath $\langle$ }\!\! \langle n_C^2  \rangle - \langle n_C \rangle^{2} 
\mbox{\boldmath $\rangle$ }\!\! _T \, .
\end{equation}
For the ground state of {\em N} independent rungs per unit volume 
$|g\rangle\!=\!\prod_j |\tilde{L}_j\rangle$, we calculate 
$g_C\!=\!N[N\!-\!(N\!-\!1)4u^2v^2\!-\!N(v^2\!-\!u^2)^2]$. Because 
$(v^2\!-\!u^2)^2\!=\!1-4u^2v^2$, it follows that $g_C\!=\!N(2uv)^2$.  Therefore, 
$g_C\!=\!0$ if the two sites on a rung are completely independent: 
$u\!=\!1$ and $v\!=\!0$, for $t_{\bot}$=0 and $\Delta\!\neq\!0$. On the other hand, 
$g_C$ has its maximum value $g_C\!=\!N$ when the rungs form homo-polar molecules: 
$u\!=\!v\!=\!1/\sqrt{2}$, for $t_{\bot}\!\neq\!0$ and $\Delta\!=\!0$. Noticing that
$1/(2uv)^2\!=\!1+(\Delta/2t_{\bot})^2$ one can see that Eq.\ \ref{sigCTgc} and
Eq.\ \ref{sigCT} correspond exactly to the same result.

In performing a similar quantitative analysis for the spin-flip excitations in the single 
two-leg ladder, we have to consider the elementary excitations of the 1D S=1/2  
HB AF, system which does not have long range magnetic order at any temperature due to its 
1D character. It has been shown by Faddeev and Takhtajan\cite{faddeev} that the true elementary 
excitations of the 1D HB AF are doublets of S=1/2 spinon excitations. Each spinon is described 
by the dispersion relation:
\begin{equation}
 \epsilon^{\prime}(k)=\frac{\pi}{2}J  \sin\!k  \,  , \hspace{2cm}  0\leq k \leq\pi \, .
\label{ft1}
\end{equation}
The two-spinon continuum, with total spin S=1 or S=0, is defined by 
$\epsilon^{\prime}(q)\!=\!\epsilon^{\prime}(k_1)\!+\!\epsilon^{\prime}(k_2)$, where 
$q\!=\!k_1\!\!+\!k_2$, and $k_1$ and $k_2$ are the momenta of the spinons. 
The lower boundary is found for $k_1\!=\!0$ 
and $k_2\!=\!q$ (or the other way around): $\epsilon^{\prime}_1(q)\!=\!\frac{\pi}{2}J \sin\!q$. 
The upper one for $k_1\!=\!k_2\!=\!q/2$\,: $\epsilon^{\prime}_2(q)\!=\!\pi J  \sin(q/2)$.

In the present case, we are dealing with double spin flips generated by the 
interaction Hamiltonian $H_S$. What is then relevant is the four-spinon continuum defined by 
$\epsilon^{\prime}(q)\!=\! \sum_{i} \epsilon^{\prime}(k_i)$, where $q\!=\! \sum_{i} {k_i}$, and 
the index {\em i}  labels the four spinons. Of this excitation spectrum we would actually 
probe in an optical experiment only the $q\!=\!0$ (or equivalently $q\!=\!2\pi$) states, with 
total spin S=0. These states form a continuum at $q\!=\!0$  extending from 
$\epsilon^{\prime}_1(0)\!=\!0$ to $\epsilon^{\prime}_2(0)\!=\!2\pi J$. In fact, because 
Eq.\ \ref{ft1} has its maximum $\epsilon^{\prime}_{max}\!=\!\pi J/2$ for $k\!=\! \pi /2$, 
four spinons with the same $k\!=\! \pi /2$  correspond to the $q\!=\!2\pi$ ($q\!=\!0$)
excited state with the highest possible energy:  
$\epsilon^{\prime}_2(0)\!=\!4 \epsilon^{\prime}_{max}\!=\!2 \pi J$.

If we now go back to Eq.\ \ref{sigS}, we see that the evaluation of the integrated optical 
conductivity for the spin excitations is problematic because it requires a detailed analysis 
of the matrix elements for the states within the four-spinon continuum, at $q\!=\!0$.  
We can simplify the problem by means of a very crude approximation and obtain a result which, 
although not rigorous, yet is meaningful for a comparison with the experiment. Let us  
replace in Eq.\ \ref{sigS} the quantity ($E_n\!-\!E_g$) with the average energy value of the 
four spinon continuum: $E_{S}\!=\!\pi J_{\parallel}$. In this way, as in the case of the CT 
excitations, we can write:
\begin{equation}
\int_{S} \sigma_{1}(\omega) d\omega =  \frac {\pi q_m^2 d_{\bot}^2} {4 \hbar^{2}V} 
  E_{S} \, g_S(T) \,  , 
\label{sigSgc}
\end{equation}
where $g_S(T)$ is the spin-correlation function defined as: 
\begin{equation}
g_S(T)= 4 \mbox{\boldmath $\langle$ }\!\! \langle h_S^2  \rangle - \langle h_S \rangle^{2} 
\mbox{\boldmath $\rangle$ }\!\! _T \, .
\label{spincf}
\end{equation}

One has to note that a non zero value of $g_S$ requires that the total Hamiltonian $H_T$ 
and the interaction Hamiltonian $H_S$ do not commute with each other. This condition would 
be realized including, in $H_T$ or in $h_S$, {\em nnn} interaction terms which would, unfortunately, 
complicate the problem considerably.  However, a finite value of $g_S$ is obtained also 
in case of an AF broken symmetry of the ground state. We can then estimate an upper limit for $g_S$  
starting from a long range N\'{e}el ordered state, i.e.  
$|g\rangle\!=\!\prod_j |\tilde{L}_{2j,\uparrow} \tilde{L}_{2j+1,\downarrow}\rangle$.  Over 
this ground state we can calculate $g_S\!=\!N$. On the other hand, for a random orientation 
of spins we would obtain $g_S\!=\!0$. The real spin configuration of the system is of course 
something in between these two extreme cases and the value of $g_S$ depends on the details of 
the many-body wave function of the spins. In particular, it depends on the probability of having 
fragments of three neighboring spins ordered antiferromagnetically. Therefore, $g_S$ is higher 
the lower is the temperature and reaches for $T\!\rightarrow\!0$ its maximum value, which would be 
 strictly smaller than $g_S\!=\!N$ in a truly 1D system. In relation to the 
interpretation of the optical conductivity data of $\alpha^{\prime}$-NaV$_2$O$_5$  
within the charged magnon model, it is important to note that, in a dimerized singlet 
(or triplet) configuration of the single two-leg ladder, we have $g_S\!=\!0$.

In conclusion, the intensity of the spin fluctuations relative to the CT excitations in terms 
of effective charges, for the single two-leg ladder system, is given by:
\begin{equation}
 \frac{\int_{\;S\,\;}\sigma_{1}(\omega)d\omega}{\int_{CT}\sigma_{1}(\omega)d\omega}
 = \frac{q_m^2g_{S} E_{S}} {q_e^2 g_{C} E_{CT}} \,  .
\label{sigCT/S} 
\end{equation}
On the basis of the analysis presented above, we can state that the maximum limiting 
value for this quantity, in the limit $U\!\rightarrow\!\infty$ and with 
$J_{\parallel}$ given by Eq.\ \ref{jpara}, is:
\begin{equation}
 \frac{\int_{\;S\,\;}\sigma_{1}(\omega)d\omega}{\int_{CT}\sigma_{1}(\omega)d\omega}
 \leq \frac {9\pi}{32} \frac {\Delta^2 J_{\parallel}^4}{t_{\parallel}^2 \, t_{\bot}^4}\, .
\label{lsigCT/S} 
\end{equation}

\section{Optical Spectroscopy}
\label{nvoptsr}

We investigated the optical properties of $\alpha^{\prime}$-NaV$_2$O$_5$ in the frequency range
going from 30 to 32\,000 cm$^{-1}$. High-quality single crystals were grown by high 
temperature solution growth from a vanadate mixture flux.\cite{ueda}  
The crystals, with dimensions of 1, 3, and 0.3 mm along the {\em a}, {\em b}, and {\em c} axes, 
respectively, were aligned by conventional 
Laue diffraction, and mounted in a liquid He flow cryostat to study the temperature dependence of 
the optical properties between 4 and 300 K. Reflectivity measurements, in near normal incidence 
configuration ($\theta\!\sim\!10^{\circ}$), were performed on two different Fourier transform 
spectrometers: A Bruker IFS 113v, in the frequency range going 20\,-\,7000 cm$^{-1}$, and 
a Bomem DA3, between 6000 and 32\,000 cm$^{-1}$. Polarized light was used, in order to probe the 
optical response of the crystals along the {\em a} and {\em b} axes. 
The absolute reflectivity was obtained by calibrating the data acquired 
on the samples against a gold mirror, from low frequencies up to 15\,000 cm$^{-1}$, and an aluminum 
mirror for higher frequencies. The optical conductivity was calculated from the reflectivity data 
using Kramers-Kronig relations.
\begin{figure}[t]
\centerline{\epsfig{figure=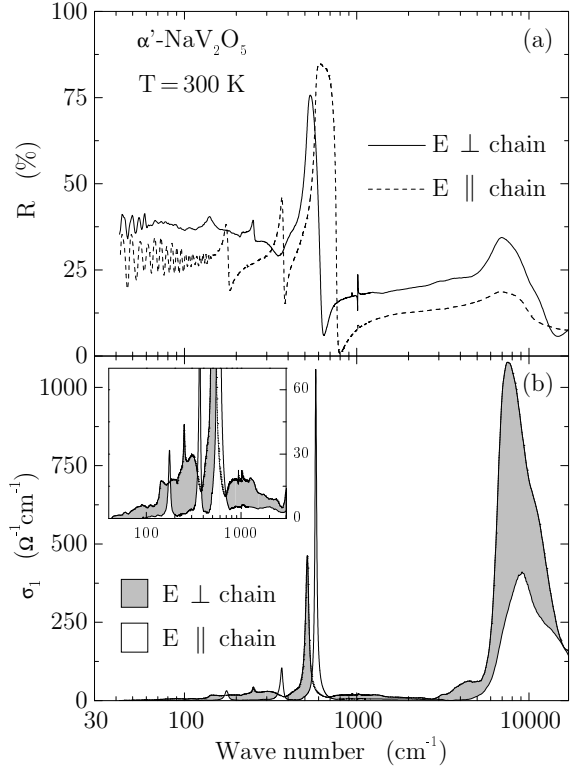,width=7.5cm,clip=}}
\vspace{0.3cm}
\caption{Optical spectra of $\alpha^{\prime}$-NaV$_2$O$_5$ at 300 K for 
$E\!\parallel\!a$ (i.e., $\perp$ to the chain direction) and $E\!\parallel\!b$ 
(i.e., $\parallel$ the chain direction), in the frequency range going from 30 to 17\,000 cm$^{-1}$.
Panel (a), and (b) show the results for reflectivity, and optical conductivity, respectively. 
In the inset of panel (b) an enlarged  view of $\sigma_{1}(\omega)$ 
from 40 to 3000 cm$^{-1}$ is presented.}
\label{nvab300}
\end{figure}

Let us now, as an introduction to what we will discuss in detail in the following sections, 
describe briefly the main features of the optical spectra of $\alpha^{\prime}$-NaV$_2$O$_5$, over the entire frequency 
range we covered  with our experimental systems. In Fig.\ \ref{nvab300} we present reflectivity 
and conductivity data of $\alpha^{\prime}$-NaV$_2$O$_5$ at 300 K for  $E\!\parallel\!a$ \,(i.e., $\perp$ to the chain 
direction) and $E\!\parallel\!b$  \,(i.e., $\parallel$ the chain direction), in the frequency 
range going from 30 to 17\,000 cm$^{-1}$. We can observe a very rich and peculiar excitation 
spectrum, characterized by strong anisotropy for direction parallel and perpendicular to the 
V-O chains. 

For $E\!\parallel$\,chain, the far-infrared region ($\omega\!<$\,800 cm$^{-1}$) is 
characterized by strong optical phonon modes, with no background conductivity, as expected for 
an ionic insulator. At higher frequencies a strong electronic absorption (peaking  
at $\sim\!9000$ cm$^{-1}$) is directly observable in both reflectivity and conductivity. 
Moreover, in the mid-infrared region (between 800 and 5000 cm$^{-1}$) we detected a very weak 
continuum of excitations, as shown by the finite value of the conductivity (see inset of Fig.\ 
\ref{nvab300}b) and by the reflectivity which is continuously raising in that frequency range 
(Fig.\ \ref{nvab300}a). 

On the other hand, for $E\!\perp$\,chain, a weak broad band of optical absorption, additional to 
the phonon modes, was detected in the far-infrared region. This is directly recognizable in the 
reflectivity spectrum (Fig.\ \ref{nvab300}a): Reducing the frequency of the incoming light, 
contrary to what observed for $E\!\parallel$\,chain, we can observe an overall increase in 
reflectivity which masks the contribution due to the sharp phonon lines. At the same time, the 
interference fringes, due to Fabry-Perot resonances, which were dominating the optical 
reflectivity below 140 cm$^{-1}$ for $E\!\parallel$\,chain, are almost completely suppressed for 
$E\!\perp$\,chain, indicating a stronger absorption for the latter polarization. In the optical 
conductivity, a continuum, with a broad maximum in the far infrared (see inset of Fig.\ 
\ref{nvab300}b), extends from very low frequencies up to the electronic excitation at 
$\sim\!7500$ cm$^{-1}$\,$\sim\!1$\,eV, which is now much more intense than the similar one 
detected for $E\!\parallel$\,chain. Moreover, along the direction perpendicular to the V-O chains, 
we can observe an additional shoulder in the conductivity spectrum, between 3000 and 5000 cm$^{-1}$. 

Finally, for both polarizations, an absorption edge due to the onset of 
charge-transfer transitions was detected at $\sim$\,3\,eV (not shown), in agreement with  
Ref.\,\onlinecite{golubchik}.

\subsection{Phonon Spectrum: High Temperature Phase}
\label{nvhtphon}

\begin{figure}[t]
\centerline{\epsfig{figure=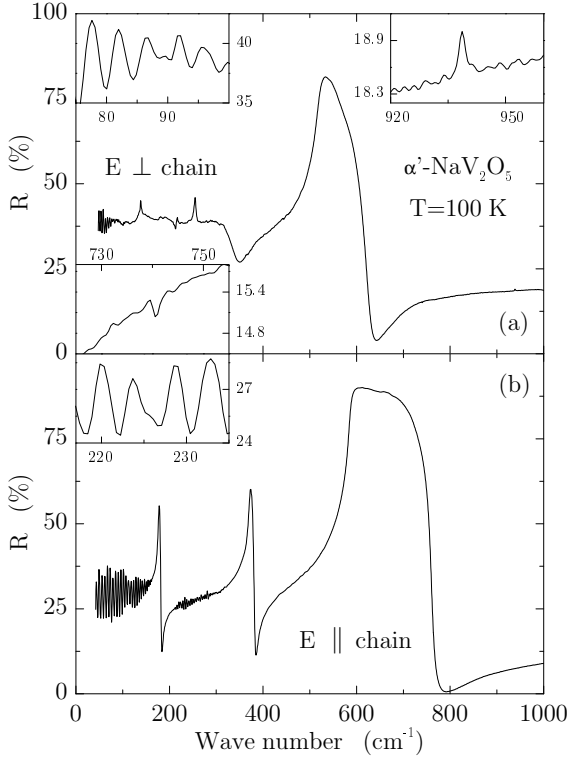,width=7.5cm,clip=}}
\vspace{0.3cm}
\caption{Reflectivity of $\alpha^{\prime}$-NaV$_2$O$_5$ in the undistorted 
phase at 100 K. The spectra are shown for $E$ perpendicular and 
parallel to the chain direction in panel (a) and (b), respectively. An enlarged view of the 
frequency regions, where very weak optical phonons were detected, is given in the insets.}
\label{nvab100r}
\end{figure}

We will now concentrate on the phonon spectra of the high temperature undistorted 
phase of $\alpha^{\prime}$-NaV$_2$O$_5$. We will try to assess the symmetry issue by comparing 
the number and the symmetry of the experimentally observed optical vibrations to the results 
obtained from the group theoretical analysis for the two different space groups proposed, for 
$\alpha^{\prime}$-NaV$_2$O$_5$, in the undistorted phase. In Fig.\ \ref{nvab100r} we present the 
reflectivity data for {\em E} perpendicular and parallel to the V-O chains (i.e., along the 
{\em a} and {\em b} axes, respectively), up to 1000 cm$^{-1}$, 
which covers the full phonon spectrum for these two crystal axes. In this respect, 
different is the case of the {\em c} axis: Along this direction a phonon mode was detected 
at $\sim$\,1000 cm$^{-1}$, as we will show in section\ \ref{nvsld} while discussing 
the results reported in Fig.\ \ref{nvabplrz}. In Fig.\ \ref{nvab100r} the data are shown for 
$T\!=\!100$ K, temperature which is low enough for the phonons to be sufficiently sharp and 
therefore more easily detectable. At the same time, it is far from $T_{\rm{c}}$, so that we can 
identify the phonon spectrum of the undistorted phase without having any contribution 
from the one of the low temperature distorted phase.

In Fig.\ \ref{nvab100r} three pronounced phonon modes are 
observable for $E\!\parallel$\,chain 
($\omega_{\rm{TO}}$\,$\approx$\,177, 371, and 586 cm$^{-1}$, at 100 K), and three for 
$E\!\perp$\,chain ($\omega_{\rm{TO}}$\,$\approx$\,137, 256, and 518 cm$^{-1}$, at 100 K). 
It has to be mentioned that the feature at 213 cm$^{-1}$  along the {\em a} axis, which looks like  
an antiresonance, is a leakage of a {\em c}-axis mode\cite{smirnov} due to the finite angle of 
incidence of the radiation on the sample ($\theta\!\sim\!10^{\circ}$), and to the use of 
{\em p}-polarization (therefore with a finite $E\|c$ component) to probe the {\em a}-axis 
optical response. Analyzing in detail the spectra few more lines are 
observable, as shown in the insets of  Fig.\ \ref{nvab100r}. Along the {\em b} axis a fourth 
phonon is present at 225 cm$^{-1}$: Being the sample a 300 $\mu$m thick platelet characterized 
by some transparency in this frequency region, Fabry-Perot interferences with a sinusoidal pattern
are measurable. At 225 cm$^{-1}$ the pattern is not regular anymore, indicating the 
presence of an absorption process. From the temperature
dependence, it can be assigned to a lattice vibration. Similarly, three more phonons 
are detected along the {\em a} axis at 90, 740 and 938 cm$^{-1}$. 
\begin{figure}[t]
\centerline{\epsfig{figure=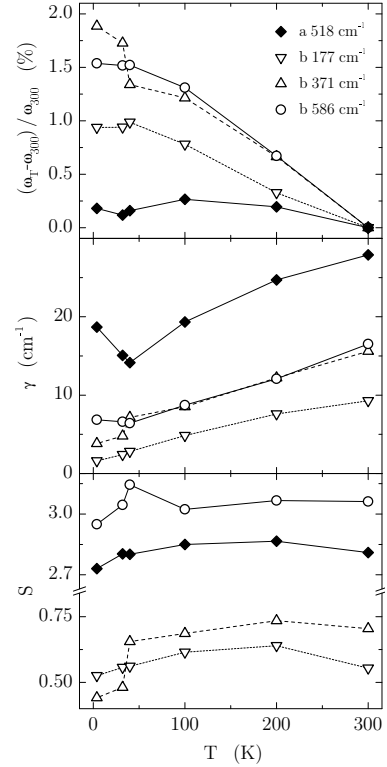,width=4.95cm,clip=}}
\vspace{0.2cm}
\caption{Temperature dependence of resonant frequency shift, damping, and oscillator strength 
(obtained by fitting the data with Fano line shapes), 
for the most intense optical phonons of the undistorted phase of $\alpha^{\prime}$-NaV$_2$O$_5$.}
\label{nvpara}
\end{figure}
%

%
\begin{table}[b]
\vspace{-0.2 cm}
\begin{center}   
\begin{tabular}{c|c|rrrrrr} 
Pol. & T\,(K) & \multicolumn{6}{c}{$\omega_{\rm{TO}}$  (cm$^{-1}$)}   \\   \hline  
\vspace{-.2cm}     & 4   &  91.12 & 139.15 & 255.49 & 517.30 & 740.62 & 938.21   \\ 
\vspace{-.2cm} $E\!\parallel\!a$  & & & & & & &   \\  
             & 100 &  89.67 & 136.81 & 255.82 & 517.74 & 740.49 & 938.49  \\  \hline 
 \vspace{-.2cm}           & 4   & 177.46 & 224.67 & 373.91  & 587.20  &   &   \\ 
 \vspace{-.2cm}$E\!\parallel\!b$ & & & & & & &   \\  
             &  100 & 177.18 & 225.20 & 371.02  & 585.88 &  &  \\  
\end{tabular}
\end{center}
\caption{Resonant frequencies of the optical vibrational modes characteristic of the 
undistorted phase of $\alpha^{\prime}$-NaV$_2$O$_5$, obtained by fitting the optical 
phonons with Fano line shapes.}
\label{nvtabphon}
\end{table}  

In conclusion, 6 {\em a}-axis 
and 4 {\em b}-axis phonons were detected on $\alpha^{\prime}$-NaV$_2$O$_5$, in the high temperature undistorted phase. The 
resonant frequencies of these vibrational modes are summarized, for two different temperatures, 
in table\ \ref{nvtabphon}. These results compare better with the group theoretical 
analysis for the centrosymmetric space group  {\em Pmmn} (see Eq.\ \ref{nvrepme}), which gave 7 
and 4 vibrational modes for ${\rm{E}}\|a$ and ${\rm{E}}\|b$, respectively, than with the analysis 
for the noncentrosymmetric $P2_1mn$. In fact, in the latter case 15 and 7 phonons ought to be 
expected (see Eq.\ \ref{nvrepcg}). 
Therefore, in agreement with the x-ray structure redetermination,\cite{meetsma,schnering,smolinski} 
also the optical investigation of the phonon spectra indicates the centrosymmetric space group 
{\em Pmmn} as the most probable one for $\alpha^{\prime}$-NaV$_2$O$_5$, at temperatures above  
$T_{\rm{c}}$. 

However, $P2_1mn$ is in principle not completely ruled out as a possible space group on the 
basis of the optical phonon spectra. 
In fact, group theory gives information only  about the number of phonons that one could expect 
but not about their resonant frequencies and, in particular, their effective strength which, in 
the end, determines whether a mode is detectable or not. The only really conclusive answer would 
have been to detect more modes than allowed by symmetry for the space group {\em Pmmn}. In 
the present situation, it may be possible that the correct space group is still the 
non-centrosymmetric 
$P2_1mn$ and that some of the phonons have escaped detection because the frequency is 
too low for our setup or due to a vanishing oscillator strength. In this respect, we have to 
stress that, indeed, the oscillator strength of the additional modes expected for the 
noncentrosymmetric $P2_1mn$ would have to be small because, as indicated from the x-ray diffraction 
analysis,\cite{meetsma} if a deviation from centrosymmetry is present, it is smaller than 0.03 \AA. 
We have also been performing a lattice dynamical calculation for both space groups which 
shows that, removing the center of inversion and changing the valence of the V ions 
(i.e., from the uniform V$^{4.5+}$ for all sites to an equal number of V$^{4+}$ and V$^{5+}$ sites), 
the number of modes characterized by a finite value of the oscillator strength does not increase, 
even if a strictly long range charge ordering in 1D chains of V$^{4+}$ and of V$^{5+}$ 
ions is assumed.

It has to be mentioned that many independent investigations of the Raman 
and infrared phonon spectra, on $\alpha^{\prime}$-NaV$_2$O$_5$, were 
reported.\cite{anvprl,golubchik,smirnov,popova1,smirnovprb,popovic,fischer,lemmens,kuroe,popova2} 
Of course, there was not immediately a perfect agreement between all the different data. 
As a matter of fact, in 
the first experimental studies of lattice vibrations,\cite{golubchik,popova1} the optical spectra 
there presented were interpreted as an evidence for the noncentrosymmetric space group $P2_1mn$. 
At this stage, we believe that the results of our investigation performed on high 
quality single crystals, give a very complete picture of the infrared vibrational modes of 
$\alpha^{\prime}$-NaV$_2$O$_5$, in the undistorted phase, for $E\|a$ and $E\|b$. Moreover, our 
results are completely confirmed by those recently obtained by Smirnov 
{\em et al.}\cite{smirnov,smirnovprb} (who could measure also the {\em c}-axis phonon spectrum), 
on samples of different origin.

Before moving on to the phonon spectra characteristic of the low temperature 
distorted phase of $\alpha^{\prime}$-NaV$_2$O$_5$, we would like to discuss the temperature dependence, from 4 to 300 K, 
of the lattice vibrations which are already present in the uniform phase. In order to extract, from 
the experimental data, information about their parameters, we had to fit the data using 
Fano line shapes for the phonon peaks.\cite{fano} In fact, most of them, in particular along the 
{\em a} axis, are characterized by a strong asymmetry indicating an interaction with the 
underlaying low frequency continuum. In this case the symmetrical Lorentz line shape was not 
suitable and the asymmetrical Fano profile had to be used: The latter model contains an additional 
parameter which takes care of the degree of asymmetry of an excitation line, due to the coupling 
between that discrete state and an underlaying continuum of states.  
\begin{figure}[t]
\centerline{\epsfig{figure=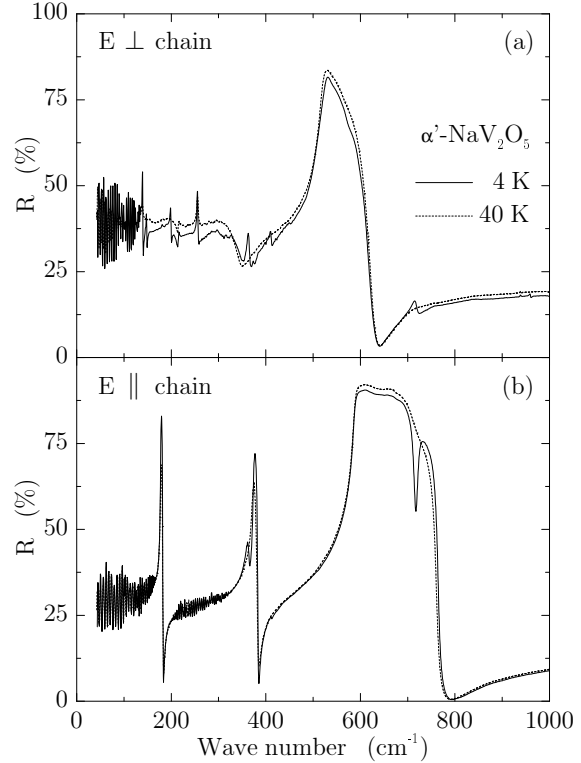,width=7.5cm,clip=}}
\vspace{0.3cm}
\caption{Reflectivity spectra of $\alpha^{\prime}$-NaV$_2$O$_5$ measured for $E\!\perp$\,chain (a) 
and $E\!\parallel$chain (b) below (4 K) and just above (40 K) the phase transition. Along both 
axes, new phonon lines activated by the phase transition are observable for 
$T\!<\!T_{\rm{c}}$.}
\label{nvab4r}
\end{figure}

%
\begin{table}[b]
\vspace{-0.2 cm}
\begin{center}   
\begin{tabular}{c|ccccccccccc}  
Pol. (LT)  & \multicolumn{11}{c}{$\omega_{\rm{TO}}$  (cm$^{-1}$)}   \\ \hline  
 $E\!\parallel\!a$   & 101 & 127 & 147 & 199 & --- & 362 & 374 & 410 & 450 & 717 & 960 \\ \hline  
 $E\!\parallel\!b$   & 102 & 128 & --- & --- & 234 &  362 & --- & 410 & --- & 718 & 960 \\ 
\end{tabular}
\end{center}
\caption{Resonant frequencies, at 4 K, of the zone boundary folded modes detected on 
$\alpha^{\prime}$-NaV$_2$O$_5$ in  the low temperature (LT) phase, with $E\!\perp$\,chain 
and  $E\!\parallel$\,chain.}
\label{nvtabphonsp}
\end{table}  

The results for the 
percentage change in resonant frequency, the damping, and the oscillator strength of the most 
intense optical phonons of the undistorted phase of $\alpha^{\prime}$-NaV$_2$O$_5$, are plotted in 
Fig.\ \ref{nvpara}, between 4 and 300 K. One can observe that some of the modes show a sudden 
change in their parameters between 40 and 32 K, i.e., upon reducing the temperature below 
$T_{\rm{c}}\!=\!34$ K. 
In particular, the oscillator strength of the phonons at 371 and 586 cm$^{-1}$, along the {\em b} 
axis, decreases across the phase transition, suggesting either a transfer of spectral weight to 
zone boundary folded modes activated by the phase transition or to electronic degrees of freedom.
We will see later, in the course of the paper, that for the chain direction the reduction of 
the phonon oscillator strength will be compensated mainly by the strength gained by the zone 
boundary folded modes activated by the phase transition. On the other hand, for $E\!\perp$\,chain, 
the change in  spectral weight of the electronic excitations will 
appear to be particularly important.

\subsection{Phonon Spectrum: Low Temperature Phase}
\label{nvltphon}

Upon cooling the sample below $T_{\rm{c}}$=34 K, significant changes occur in the optical 
phonon spectra as one can see in Fig.\ \ref{nvab4r}, where we present the reflectivity spectra of 
$\alpha^{\prime}$-NaV$_2$O$_5$ measured, for $E\!\perp$\,chain (a) and $E\!\parallel$\,chain (b), 
below (4 K) and just above (40 K) the phase transition. Contrary to the case of CuGeO$_3$, 
where we could clearly observe in reflectivity only a single, very weak 
additional phonon,\cite{sp1,sp2,sp3}  10 new lines are  detected for $E\!\perp$\,chain 
($\omega_{\rm{TO}}$\,$\approx$\,101, 127, 147, 199, 362, 374, 410, 450, 717, and 960  cm$^{-1}$), 
and 7 for $E\!\parallel$\,chain ($\omega_{\rm{TO}}$\,$\approx$\,102, 128, 234, 362, 410, 718, 
and 960 cm$^{-1}$), some of them showing a quite large  oscillator strength. The stronger 
intensity of the lines, with respect to the case of CuGeO$_3$, suggests larger lattice 
distortions in $\alpha^{\prime}$-NaV$_2$O$_5$, on going through the phase transition or, 
alternatively, a larger electronic polarizability. The resonant 
frequencies (at $T\!\!=\!\!4$ K) of all the detected zone boundary folded modes are summarized in 
table\ \ref{nvtabphonsp}. The observation of many identical or almost identical frequencies for 
both axes, which is an important information for a full understanding of the structural 
distortion, is an intrinsic property of the low temperature phase. Experimental errors, like 
polarization leakage or sample misalignment, are excluded from the well-defined anisotropy of the  
phonon spectrum of the undistorted phase. A more detailed discussion of this point will be 
presented in section \ref{nvsld}, in relation to the reflectivity data reported 
in Fig.\ \ref{nvabplrz}.
\begin{figure}[t]
\centerline{\epsfig{figure=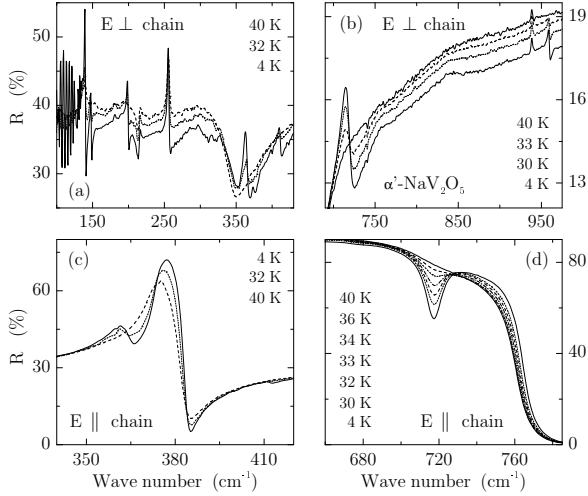,width=7.75cm,clip=}}
\vspace{0.3cm}
\caption{Detailed temperature dependence of some of the zone boundary folded modes observed 
in the reflectivity  spectra, for $T\!<\!T_{\rm{c}}$, with $E\!\perp$\,chain (a,b) and  
$E\!\parallel$\,chain (c,d).}
\label{nvabrpks}
\end{figure}
\begin{figure}[t]
\centerline{\epsfig{figure=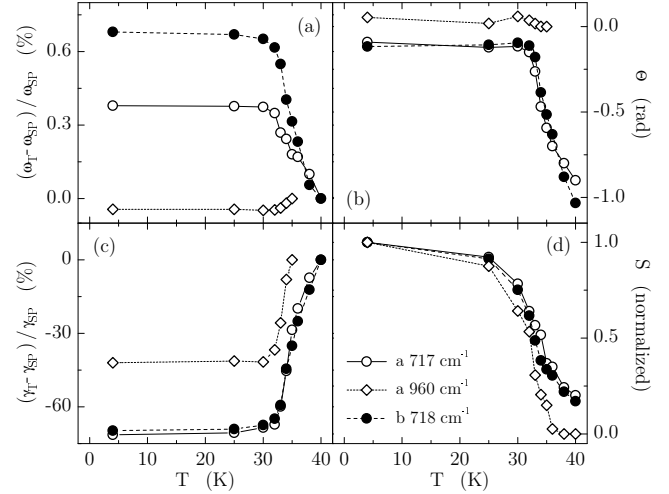,width=8.5cm,clip=}}
\vspace{0.3cm}
\caption{Temperature dependence of the change in resonant frequency (a) and 
damping (c), of the normalized oscillator strength (d), and of the Fano asymmetry parameter 
(b), for the folded phonons observed below the phase transition at 717 and 960 cm$^{-1}$  
with $E\!\parallel\!a$, and at 718 cm$^{-1}$ with $E\!\parallel\!b$.}
\label{nvparasp}
\end{figure}

An enlarged view of few of the frequency regions where zone boundary folded modes were detected, 
is given in Fig.\ \ref{nvabrpks}. We observe the gradual growing of the these modes, whose 
oscillator strength increases the more the temperature is reduced below $T_{\rm{c}}$=34 K. 
Moreover, the reflectivity data acquired for $E\!\perp$\,chain [panel (a) in Fig.\ \ref{nvab4r} 
and panels (a) and (b) in Fig.\ \ref{nvabrpks}] show that also the underlaying continuum has a 
considerable temperature dependence: The reflectivity is decreasing, over the entire 
frequency range, upon reducing the temperature from 40 to 4 K. In particular, in 
Fig.\ \ref{nvabrpks}a, we see that very pronounced interference fringes, due to Fabry-Perot 
resonances, are present at 4 K for $\omega\!<\!136$ cm$^{-1}$, indicating a 
particularly strong reduction of the absorption in that frequency region. This effect is, in our 
opinion, related to the opening of the magnetic gap in the excitation spectrum, and will be 
discussed in detail in section\ \ref{nvoc}.

Of all the detected folded modes, for experimental reasons (i.e., higher sensitivity of the 
detector and therefore higher accuracy of the results in this frequency range), we carefully 
investigated the ones at 718 ($E\!\parallel\!a,b$) and 960 cm$^{-1}$ ($E\!\parallel\!a$), 
for several temperatures between 4 and 
40 K. We fitted these phonons to a Fano profile because the 718 cm$^{-1}$ modes show an 
asymmetrical line-shape. The results of the percentage change in resonant frequency and 
damping, of the normalized oscillator strength, and of the Fano asymmetry parameter 
are plotted versus temperature in Fig.\ \ref{nvparasp}.  
We use here an asymmetry parameter $\Theta$ (rad) defined  in such 
a way that the larger the value of $\Theta$, the stronger is the asymmetry of the line: A Lorentz 
line-shape is recovered for $\Theta$=0.\cite{fesi}

Let us now start by commenting on the 
results obtained for the oscillator strength: In Fig.\ \ref{nvparasp}d we see that {\em S} has a 
similar behavior for the three different lines. However, the 960 cm$^{-1}$ peak vanishes at 
$T_{\rm{c}}$ whereas the  two 718 cm$^{-1}$ modes have still a finite intensity at $T$=40 K and, 
as a matter of fact,  disappear only  for $T\!>$\,60\,-70 K. At the same time, the line-shape of the 960 
cm$^{-1}$ mode is perfectly lorentzian at all temperatures, whereas the two other  phonons show a 
consistently increasing asymmetry for $T\!>\!32$ K. Also the increase in resonant frequency 
(Fig.\ \ref{nvparasp}a) and the reduction of the damping (Fig.\ \ref{nvparasp}c) are much more 
pronounced for the two 718 cm$^{-1}$ modes. From these results we conclude that the second-order 
character of the phase transition is nicely shown by the behavior of {\em  S} for the 960
cm$^{-1}$ folded mode. On the other hand, pretransitional fluctuations manifest themselves in the 
finite intensity of the 718 cm$^{-1}$ modes above $T_{\rm{c}}$, and in the extremely large value 
of $\Theta$ and $\gamma$ for these lines ($\gamma\!\sim\!35$ cm$^{-1}$ at 40 K). 
In other words, the lattice distortion is already taking place in the system, at temperature much 
higher than $T_{\rm{c}}$, but with a short range character, whereas a coherent long range 
distortion is realized only below the phase transition temperature. In this respect, it is worth 
mentioning that similar pretransitional fluctuations, below 70 K, have been observed also in the 
course of a study of the propagation of ultrasonic waves along the chain direction of 
$\alpha^{\prime}$-NaV$_2$O$_5$,\cite{fertey} and, very recently, in x-ray diffuse scattering 
measurements.\cite{ravy}

\subsection{Symmetry of the Lattice Distortion}
\label{nvsld}

We saw in the previous section that many of the folded zone boundary phonons, activated by 
the phase transition, show the same resonant frequency for $E\|a$ and $E\|b$ 
(see table\ \ref{nvtabphonsp}), even though they are characterized by a different oscillator 
strength along the two different axes (e.g., $S\!\sim\!0.021$ for $E\|a$ and $S\!\sim\!0.014$ for 
$E\|b$, for the 718 cm$^{-1}$ modes at 4 K). This is quite a surprising result because the phonon 
spectrum of the undistorted phase has a well defined anisotropy, at any 
temperature, with different resonant frequencies for vibrations polarized along the {\em a} 
and {\em b} axes (see table\ \ref{nvtabphon}). Similar anisotropy should then be present 
below the phase transition if the system is still orthorhombic, even for the folded modes. 
\begin{figure}[t]
\centerline{\epsfig{figure=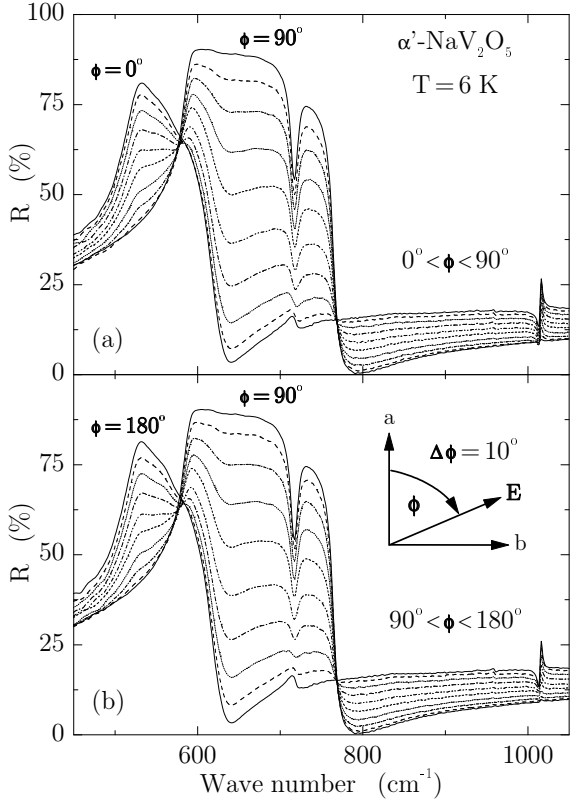,width=7.5cm,clip=}}
\vspace{0.3cm}
\caption{Reflectivity of $\alpha^{\prime}$-NaV$_2$O$_5$, at 6 K, for several orientations of the 
electric field of the light in the $a$-$b$ plane (as sketched in panel b, with 
$\Delta\phi\!=\!10^{\circ}$). The values $\phi\!=\!0^{\circ}$,180$^{\circ}$ correspond to 
$E\!\perp$\,chain, and $\phi\!=\!90^{\circ}$  to $E\!\parallel$\,chain.}
\label{nvabplrz}
\end{figure}

To further check these findings, we performed reflectivity measurements on a single crystal of 
$\alpha^{\prime}$-NaV$_2$O$_5$, at the fixed temperature of 6 K, rotating the 
polarization of the incident light of an angle $\phi$ (see Fig.\ \ref{nvabplrz}). The sample was 
aligned in such a way to have the electric field of the light parallel to the {\em a} axis of 
the crystal, for $\phi\!=\!0^{\circ}$ and $180^{\circ}$, and parallel to the {\em b} axis, 
for $\phi\!=\!90^{\circ}$. For these polarizations of the light the obtained results are, 
obviously, identical to those presented in Fig.\ \ref{nvab4r}.  However in Fig.\ \ref{nvabplrz} 
the data are shown in a different frequency range, extending up to 1050 cm$^{-1}$. We can then 
observe at $\sim\!1000$ cm$^{-1}$ along the {\em a} axis, which again was probed with 
{\em p}-polarized light, the antiresonance due to the leakage of a {\em c}-axis 
phonon.\cite{smirnov} Starting from $\phi\!=\!0^{\circ}$ and turning the polarization in steps 
$\Delta\phi\!=\!10^{\circ}$, we observe a gradual decrease of the {\em a} axis contribution and, 
at the same time, an increase of the {\em b} axis contribution to the total reflectivity. Moreover, 
the set of curves plotted in Fig.\ \ref{nvabplrz} satisfy the following relations:
\begin{equation}
           R(45^{\circ})+R(-45^{\circ})=R(0^{\circ})+R(90^{\circ}) \,\, ,  
\end{equation}
\vspace{-.85 cm}
\begin{equation}
           R(\phi)=R(0^{\circ})cos^2(\phi)+R(90^{\circ})sin^2(\phi)\,\, .
\label{Rphi}
\end{equation}
This suggests that the system, even in the distorted phase, has the optical axes along the crystal 
directions {\em a}, {\em b}, and  {\em c} (on the contrary, for a 
triclinic or a monoclinic structure there 
would be no natural choice of the optical axes and Eq.\ \ref{Rphi} would not be satisfied). 
However, the low temperature folded mode at 718 cm$^{-1}$ can be observed for all possible 
orientations of the electric field of the incident radiation (see Fig.\ \ref{nvabplrz}). 
Because identical resonance frequencies, along the {\em a} and the {\em b} axes of the crystal, 
were observed for six of the folded zone boundary modes, we cannot believe in an accidental 
coincidence. 
\begin{figure}[t]
\centerline{\epsfig{figure=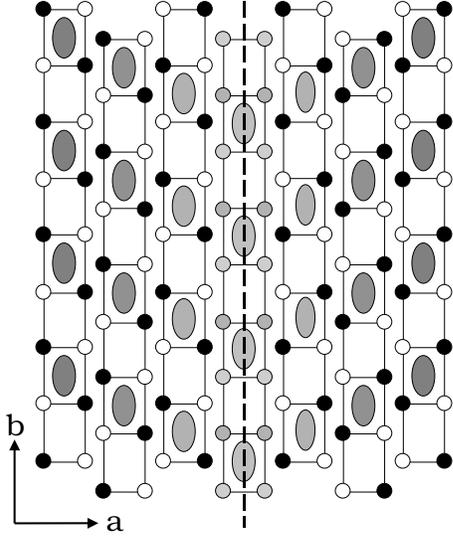,width=6cm,clip=}}
\vspace{0.3cm}
\caption{Possible charge ordering pattern in 
$\alpha^{\prime}$-NaV$_2$O$_5$, below $T_{\rm{c}}$. 
The grey ellipsoids represent the electronic charge involved in the formation 
of singlets, within the ladders, over every second plaquette formed by four V ions. On adjacent 
ladders, the charge is assumed to define an oblique charge ordering pattern 
(OCOP; see, {\em e.g.}, left-hand side).
Two different kinds of V ions are identifiable: V ions adjacent to empty plaquettes 
(white circles) and V ions adjacent to plaquettes containing a singlet (black circles), arranged 
in zig-zag patterns on each ladder. If domains with different orientation for the OCOP are 
present in the system (left and right-hand sides), 
the zig-zag ordering of the different kind of V ions is destroyed in correspondence 
of a domain wall (dashed line).}
\label{nvlowt}
\end{figure}

A possible explanation for the observed effect is as follows. 
In the low temperature phase a non-magnetic ground state is realized, and 
true singlets are present in $\alpha^{\prime}$-NaV$_2$O$_5$. The most probable situation is that 
the singlets are 
formed, within the ladders, over the plaquettes defined by four V ions. In this case, as shown in 
Fig.\ \ref{nvlowt}, within one ladder there is an alternation of `empty' plaquettes and plaquettes 
containing a singlet (represented by a grey ellipsoid in Fig.\ \ref{nvlowt}). Let us now 
assume that in adjacent ladders the charge is arranged in 
an oblique charge ordering pattern (OCOP), 
as shown in the left-hand side of Fig.\ \ref{nvlowt}. The
OCOP is consistent with the doubling of the lattice constant observed, along the 
{\em a} axis, in low temperature x-ray scattering experiments.\cite{fujii} Moreover, two different 
kinds of V ions are identifiable which could possibly explain the results obtained in low 
temperature NMR experiments\cite{ohama} (for a more detailed discussion see section\ \ref{nvdisc}). 
In fact, there are V ions adjacent to empty plaquettes (white circles) and V ions adjacent to 
plaquettes containing a singlet (black circles), arranged in zig-zag patterns on each ladder.
Note that in our picture the non-equivalence of plaquettes along the $b$-axis, 
indicated by the gray ellipsoids in Fig.\ \ref{nvlowt}, is really a 
chicken-and-egg problem: 
Either the singlet formation occurs first, causing an alternation 
along $b$ of non-equivalent plaquettes, which, in combination with the 
OCOP along $a$ gives rise to a zig-zag charge ordering. 
Alternatively, the zig-zag charge order occurs first, which, in 
combination with OCOP causes the alternation of non-equivalent plaquettes 
along $b$. The resulting charge modulations are only small 
deviations from average occupation, which we represented graphically by the 
different grey shades.

At this stage, the low temperature phonon spectra should still be characterized by a well 
defined anisotropy, with distinct eigenfrequencies for different polarizations of the electric field 
of the incoming radiation. However, there is no reason for the polarization of the folded zone 
boundary modes to be precisely along the {\em a} or the {\em b} axes. In fact, they are polarized 
with a finite angle with respect to the crystal axes (e.g., along directions parallel 
and perpendicular to the diagonal long-range singlet-pattern).  
If domains are present in the system, with opposite orientation for the 
OCOP (left and right hand sides of Fig.\ \ref{nvlowt}), each domain 
will still be characterized by a phonon spectrum with a well defined polarization, as just 
discussed. However, in presence of such domains the phonon spectra measured in the experiments
correspond to an average over many domains with the two possible orientations randomly 
distributed. As a result, almost no 
anisotropy would be found for light polarized along {\em a} and {\em b} axes.

A short wavelength alternation of descending and ascending OCOP domains 
(also a $2a\times 2b$ supercell) corresponds closely to the structure recently reported by 
L\"udecke {\em et al.}\cite{ludecke} A similar structure was 
also considered by Riera and Poilblanc as one of the many 
possible realizations of charge density wave in this system.\cite{riera}

\subsection{Optical Conductivity}
\label{nvoc}

\begin{figure}[t]
\centerline{\epsfig{figure=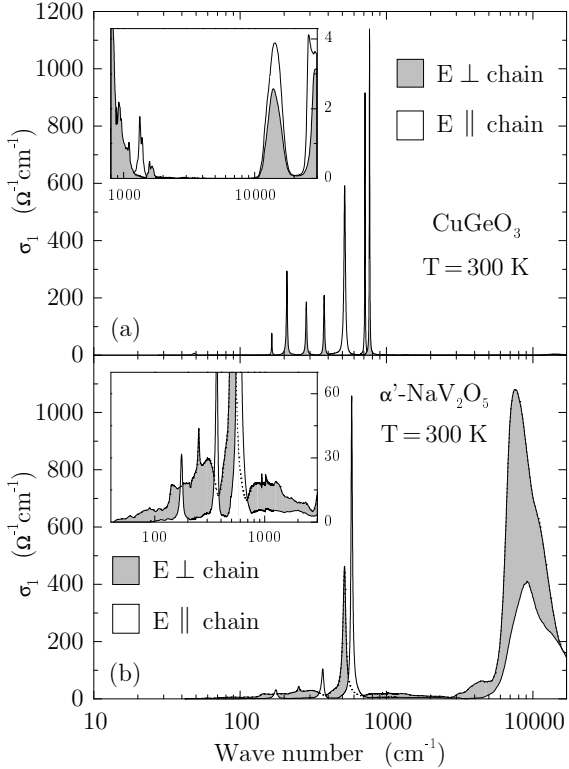,width=7.5cm,clip=}}
\vspace{0.3cm}
\caption{Optical conductivity, at 300 K, of CuGeO$_3$ (a) and $\alpha^{\prime}$-NaV$_2$O$_5$ 
(b), for $E$ parallel and perpendicular to the Cu-O or to the V-O chains. Inset of 
panel a: enlarged view of $\sigma_{1}(\omega)$ of CuGeO$_3$, from 800 to  30\,000 cm$^{-1}$.$\!$ 
Inset of panel (b):$\!$ enlarged  view of $\sigma_{1}(\omega)$ of $\alpha^{\prime}$-NaV$_2$O$_5$, 
from 40 to 3000 cm$^{-1}$.}
\label{nvcgsig}
\end{figure}

Very interesting information and a deeper understanding of the symmetry of 
$\alpha^{\prime}$-NaV$_2$O$_5$ are obtained from the detailed analysis of the optical conductivity 
data, in particular when the electronic excitations are considered and compared to the result 
relative to CuGeO$_3$ (see Fig.\ \ref{nvcgsig}). On the latter compound we observed sharp phonon 
lines below 1000 cm$^{-1}$,\cite{sp1,sp2,sp3} multiphonon 
absorption at $\sim$1500 cm$^{-1}$, very weak [note the  low values of $\sigma_1(\omega)$ in the 
inset of Fig.\ \ref{nvcgsig}a] phonon-assisted Cu {\em d-d} transitions at $\sim$\,14\,000 
cm$^{-1}$,\cite{bassi} and the onset of the Cu-O charge-transfer (CT) excitations at 
$\sim$$27\,000$ cm$^{-1}$. On $\alpha^{\prime}$-NaV$_2$O$_5$, besides the phonon lines in the far-infrared region, we detected 
features that are completely absent in CuGeO$_3$: A strong absorption peak at $\sim$8000 
cm$^{-1}$ and, in particular for $E\!\perp$\,chain, a low-frequency continuum of excitations 
(see inset of Fig.\ \ref{nvcgsig}b). 
Our goal is to interpret the complete excitation spectrum for $\alpha^{\prime}$-NaV$_2$O$_5$, trying, in this way, to 
learn more about the system on a microscopic level. In order to do that, we have first to identify  
possible candidates for the excitation processes observed in the experimental data.

\begin{figure}[t]
\centerline{\epsfig{figure=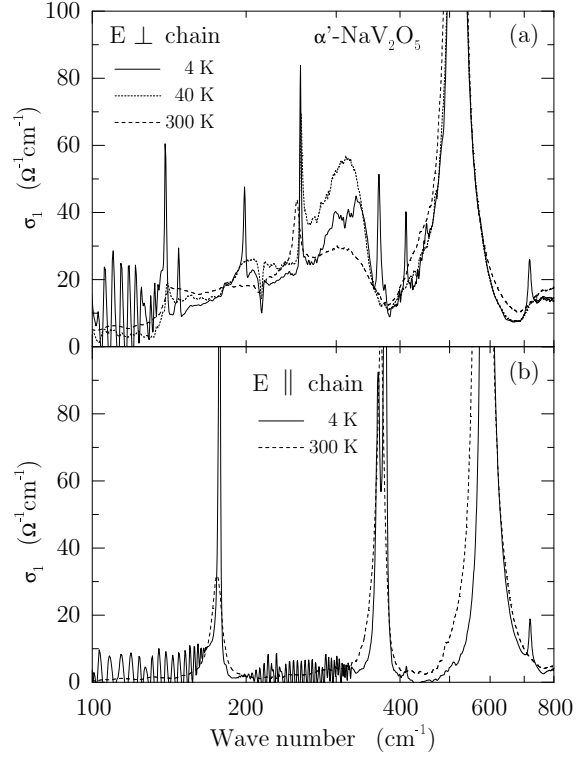,width=7.5cm,clip=}}
\vspace{0.3cm}
\caption{Far-infrared optical conductivity of $\alpha^{\prime}$-NaV$_2$O$_5$, for 
$E\!\perp$\,chain (a) and $E\!\parallel$\,chain (b). Along 
both axes new phonon lines, activated by the phase transition, are present for 
$T\!<\!T_{\rm{c}}\!=\!34$ K. Moreover, for $E\!\perp$\,chain, the temperature dependence of the 
low frequency continuum is observable.}
\label{nvab4s}
\end{figure}
\begin{figure}[t]
\centerline{\epsfig{figure=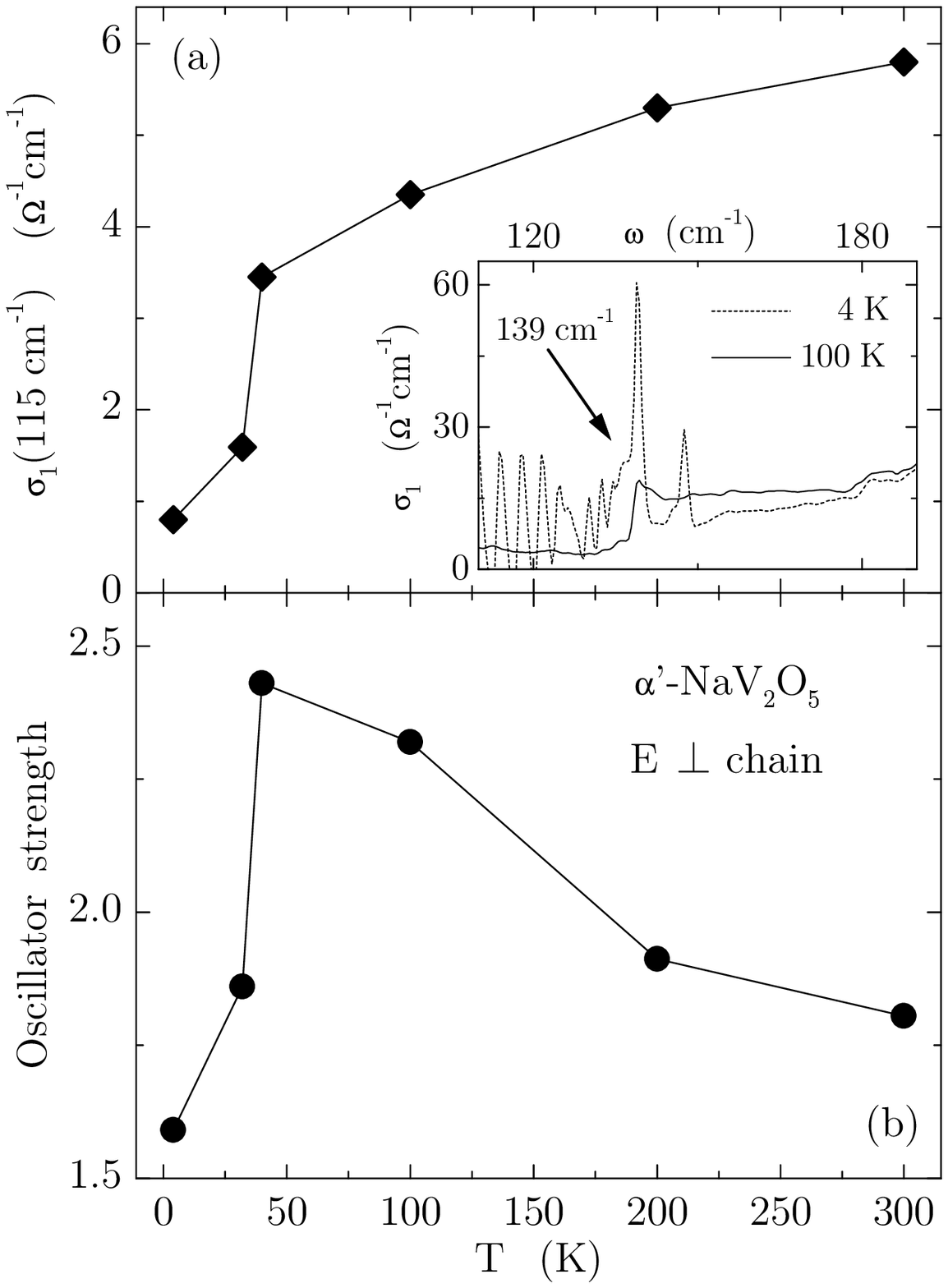,width=7.25cm,clip=}}
\vspace{0.3cm}
\caption{Panel (a): Temperature dependence of the optical conductivity at $\omega\!=\!115$ 
cm$^{-1}$, for $E\!\perp$\,chain. The values are obtained from 
Kramers-Kronig analysis, for $T\!>\!T_{\rm{c}}$, and by fitting the interference 
fringes in reflectivity, for $T\!<\!T_{\rm{c}}$. The inset shows an enlarged view of 
$\sigma_1(\omega)$, from 110 to 190 cm$^{-1}$. Panel (b): Oscillator strength of the 
low frequency electronic continuum, for $E\!\perp$\,chain, plotted versus temperature.}
\label{nvgapst}
\end{figure}

On the basis of intensity considerations, the peak at 
$\sim$8000 cm$^{-1}$\,$\sim$1\,eV, in the optical conductivity of $\alpha^{\prime}$-NaV$_2$O$_5$, has to be ascribed to an 
optically allowed excitation. Contrary to the case of the absorption detected at $\sim$\,14\,000 
cm$^{-1}$ on CuGeO$_3$, it cannot be interpreted just as a {\em d-d} exciton on the transition metal ion 
(i.e., V in the case of $\alpha^{\prime}$-NaV$_2$O$_5$). In fact, this kind of transition is in principle optically forbidden 
and only weakly allowed in the presence of a strong parity-broken crystal field or 
electron-phonon coupling. 
We have now to recall one of the main consequences of the structural analysis which suggested as 
a most probable space group, for $\alpha^{\prime}$-NaV$_2$O$_5$ in the undistorted phase, the centrosymmetric {\em Pmmn}: 
The existence of only one kind of V per unit cell, with an average 
valence of +4.5. Taking into account that the hopping parameter along 
the rungs $t_{\bot}$, is approximately a factor of two larger than the one along the legs of 
the ladder $t_{\parallel}$,\cite{horsch,nishimoto} all the other hopping amplitudes being much 
smaller (including the  nn $t_{xy}$), $\alpha^{\prime}$-NaV$_2$O$_5$ can be regarded as a 1/4-filled two-leg 
ladder system, with one V {\em d}-electron per rung shared by two V ions in a molecular bonding 
orbital.\cite{smolinski} In this context, the most obvious candidate for the strong optical 
absorption observed at 1\,eV is the on-rung bonding-antibonding transition, with a characteristic 
energy given by $2t_{\bot}$. However, on the basis of LMTO band structure calculation\cite{horsch} 
and exact diagonalization calculation applied to finite size clusters,\cite{nishimoto} 
$t_{\bot}$ was estimated to be $\sim$\,0.30-0.35\,eV, which gives a value for the  
bonding-antibonding transition lower than the experimentally observed 1\,eV. This energy mismatch, 
and the strong dependence of the intensity of the 1\,eV peak on the polarization of the 
light, are important pieces of information and will be discussed further in section\,\ref{nvcm}. 

Let us now turn to the infrared continuum detected for $E\!\perp$\,chain (see inset of 
Fig.\ \ref{nvcgsig}b). We already showed , discussing the low temperature reflectivity data (see 
Fig.\ \ref{nvab4r} and\ \ref{nvabrpks}), that this broad band of optical absorption has a 
considerable temperature dependence: The overall reflectivity was decreasing upon reducing the 
temperature and, for $T\!<\!T_{\rm{c}}$, Fabry-Perot resonances appear at 
frequencies lower than 136 cm$^{-1}$, indicating a particularly strong reduction of the optical  
absorption in that frequency region. These effects become even more evident when we consider 
the low temperature optical conductivity calculated from the reflectivity data via Kramers-Kronig 
transformations (see Fig.\ \ref{nvab4s}). However, we have to keep in mind that performing this kind of 
calculation, we will obtain unphysical oscillations for the optical conductivity in the 
frequency regions where reflectivity is dominated by interference fringes. 

In Fig.\ \ref{nvab4s}b 
we can observe, for $E\!\parallel$\,chain, only the sharpening of the 
phonon lines and the appearance of the zone boundary folded modes. On the other hand, for 
$E\!\perp$\,chain we can observe, in addition to the folded modes below 
$T_{\rm{c}}$, 
first the increase of the continuum intensity upon cooling the sample from 300 to 40 K and, 
subsequently, a reduction of it from 40 to 4 K. In the frequency region below the 139 cm$^{-1}$ 
phonon, whereas a reduction is observable in going from 300 to 40 K, not much can be said about 
the absolute conductivity of the 4 K data, because of the interference fringes. 

In order to evaluate the low temperature values of conductivity
more precisely, we can fit the interference fringes directly in reflectivity. In fact, fitting the 
period and the amplitude of them, we have enough parameters to calculate right away the real and 
imaginary part of the index of refraction and, eventually, the dynamical conductivity 
$\sigma_1$. 
The so obtained values are plotted versus the temperature in Fig.\ \ref{nvgapst}a. The data for 
100, 200, and 300 K were obtained via Kramers-Kronig transformations, whereas those for 4 and 
32 K from 
the fitting of the interferences in the range 105\,-125 cm$^{-1}$. The 40 K point, as at that 
temperature very shallow fringes start to be observable, was calculated with both methods, which 
show a rather good agreement in this case. As a results we can observe (see Fig.\ \ref{nvgapst}a) 
a pronounced reduction of $\sigma_1$, across the phase transition, corresponding to the  opening 
of a gap in the optical conductivity. 

It is difficult to evaluate precisely the gap size because 
of the phonon at 139 cm$^{-1}$ (inset of Fig.\ \ref{nvgapst}a). However, we can estimate a 
lower limit  for the gap of $\sim\!136$ cm$^{-1}$ (17$\pm$3 meV). The important 
information is that the value of the gap in the optical conductivity is approximately 
twice the spin gap value observed in inelastic neutron scattering experiments\cite{fujii} and 
magnetic susceptibility measurements,\cite{weiden} for T$<$\,34 K. This finding indicates that the 
infrared broad band of optical absorption has to be related to electronic degrees of freedom:  
Because the range of frequency coincides with the low-energy-scale spin excitations, 
the most likely candidates for this continuum are excitations involving two spin flips. 

In relation to what we just discussed, it is interesting to note the effect of the 
opening of a gap, in the optical conductivity, on the phonon located at 139 cm$^{-1}$. This can 
be clearly identified in the inset of  Fig.\ \ref{nvgapst}a, where an enlarged view of 
$\sigma_1(\omega)$ 
at low frequency is given, for 100 and 4 K. At 100 K this mode is characterized by a very 
pronounced anomaly of its line shape. As a matter of fact, fitting this phonon with a Fano line 
shape we obtained for the asymmetry parameter the value $\Theta\!=\!-1.7$ rad, indicating a  
strong coupling to the continuum. However at 4 K, once the gap is open and the intensity of the 
underlying continuum has decreased, a sharp and rather symmetrical line shape is recovered.

An additional information we can extract from the conductivity spectra is the value of the 
integrated intensity, and its temperature dependence, for the infrared broad band of optical 
absorption detected for $E\!\perp$\,chain. The oscillator strength, obtained by 
integrating $\sigma_{1}(\omega)$ (with phonons subtracted) up to $\sim$\,800 cm$^{-1}$, is 
displayed in Fig.\ \ref{nvgapst}b. It increases upon cooling down the sample from room 
temperature to $T_{\rm{c}}$, and rapidly decreases for $T\!<\!T_{\rm{c}}$. The possible 
meaning of this behavior  will be further discussed in the next section.  

As a last remark, it is important to mention that low temperature measurements of the dielectric 
constant, in the microwave frequency range, were recently performed on 
$\alpha^{\prime}$-NaV$_2$O$_5$ by different 
groups.\cite{sekine,smirnov5} The reported results show no anomaly, across the phase transition, 
for the real part of the dielectric constant along the {\em b} axis. On the other hand, along the 
{\em a} axis, a pronounced decrease below $T_{\rm{c}}$ was found. 
The observed behavior, as a matter of fact, is in very good agreement with the temperature 
dependence of the oscillator strength of the low frequency continuum present, in the optical 
spectra, along the {\em a} axis (Fig.\ \ref{nvgapst}b). This comparison is 
meaningful because the contribution of all possible optical excitations to the static 
dielectric constant is just given by their oscillator strengths. The very good agreement of 
the different results proves, as already suggested by the authors of Ref.\ \onlinecite{sekine} 
and\ \onlinecite{smirnov5}, that the change in the very low frequency dielectric constant is 
related to electronic (magnetic) degrees of freedom.

\subsection{Charged Bi-Magnons}
\label{nvcm}

In the previous section we identified possible candidates for the 1\,eV excitation (on-rung 
bonding-antibonding transition), and for the low frequency continuum detected on 
$\alpha^{\prime}$-NaV$_2$O$_5$ for $E\!\perp$\,chain (double spin flips). However, we also 
stressed that, for a symmetrical 1/4\,-filled two leg ladder system, the bonding-antibonding 
transition would be at $2t_\perp\!\sim\!0.7$\,eV 
and not at the experimentally observed value of 1\,eV. Moreover, also intensity and  polarization 
($E\!\perp$\,chain) of the continuum cannot be understood assuming the complete equivalence 
of the V sites required by the space group {\em Pmmn}. In fact, because of the high symmetry, 
no dipole moment perpendicular to the V-O chains would be associated with a double spin flip 
process. Therefore, no optical activity involving magnetic degrees of freedom would be detectable. 
Not even phonon-assisted spin excitations considered by 
Lorenzana and Sawatzky\cite{lorenzana1,lorenzana2,lorenzana3} (where the involved phonon is 
effectively lowering the symmetry of the system) could be a possible explanation for the 
observed spectra: The energy is too low for this kind of processes which, on the other hand, 
are probably responsible for the weak mid-infrared continuum in the $E\!\parallel$\,chain 
spectra (see Fig.\ \ref{nvcgsig}b and, in particular, its inset). 

\begin{figure}[t]
\centerline{\epsfig{figure=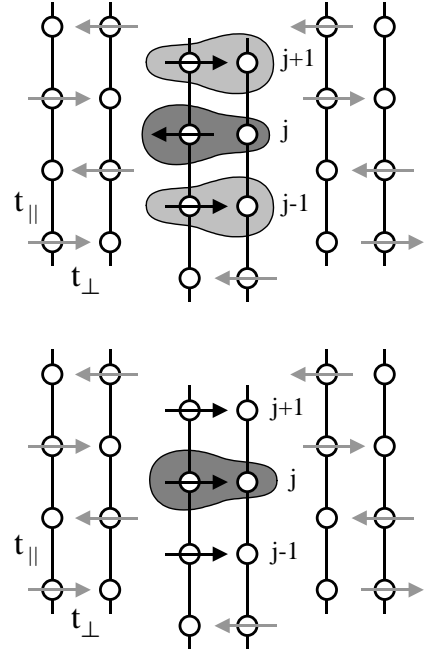,width=5.5cm,clip=}}
\vspace{0.3cm}
\caption{Pictorial description of the electric dipole moment associated with a single 
spin-flip in $\alpha^{\prime}$-NaV$_2$O$_5$, when a local breaking of symmetry in the charge  
distribution on each rung is assumed.  If three neighboring antiparallel spins are laying on the 
same leg of the ladder (top), the charge of the  $j$\,th electron is asymmetrically distributed 
not only over the $j$\,th rung but also on the nn rungs (with opposite asymmetry), because of the 
virtual hopping $t_\parallel$. If we flip its spin (bottom), the $j$\,th electron is confined to 
the  $j$\,th rung, resulting in a reduced charge density on the right leg of the ladder. 
Therefore, there is a net dipole displacement (along the rung direction) of the antiparallel 
spin configuration compared to the parallel one.}
\label{nvsflip}
\end{figure}

A possible solution is to assume a charge disproportionation on each rung of the ladders without, 
however, any particular long range charge ordering (see Fig.\ \ref{nvsflip}). The latter 
is, in fact, most probably excluded by the x-ray diffraction results and by the phonon spectra of 
 the undistorted phase. It is worth mentioning that the assumption 
of charge disproportionation has been confirmed by calculations of the optical 
conductivity by exact-diagonalization technique on finite size clusters.\cite{nishimoto1} Only in 
this way it was possible to reproduce the energy position of the 1\,eV peak, and the anisotropy 
between {\em a} and {\em b} axes (see Fig.\ \ref{nvcgsig}). We can now 
try to interpret the optical data on $\alpha^{\prime}$-NaV$_2$O$_5$ in the framework of the 
charged-magnon model developed in section\ \ref{chmgmod}.

Let us start from the  analysis of the absorption band at 1\,eV. We can model the {\em j}\,th rung, 
formed by two V ions, with the Hamiltonian $H_{j0}+H_{j\bot}$, as in Eq.\ \ref{h0} 
and\ \ref{hperp}, where the potential energy difference between the two sites $\Delta$ has the 
role of a charge disproportionation parameter: For the symmetric ladder $\Delta\!=\!0$. The 
splitting between the two bonding and antibonding eigenstates $|\tilde{L}\rangle$ and 
$|\tilde{R}\rangle$, given by $E_{CT}\!=\!\sqrt{\Delta^2+4t_{\bot}^2}$, corresponds to the 
photon {\em energy} of the optical absorption. We see that, if $\Delta\!\neq\!0$, the 
experimental value of 1\,eV can be reproduced. 

A second crucial piece of information is provided by the {\em intensity} of the absorption. 
We can calculate the integrated intensity of the 1\,eV peak, in the conductivity data, and then 
take advantage of its functional expression derived in section\ \ref{chmgmod} 
(Eq.\ \ref{sigCT}). This way we obtain from the
spectra  $|t_{\bot}|\!\approx 0.3$\,eV, value which is in very good 
agreement with those obtained from LMTO band structure calculation,\cite{horsch} and exact 
diagonalization calculation applied to finite size clusters.\cite{nishimoto} Combining this number 
with $E_{CT}\!=\!1$\,eV, we obtain $\Delta \!\approx\! 0.8$\,eV. The corresponding two eigenstates 
have 90\% and 10\% character on either side of the rung. Therefore, the valence of the two 
V-ions is 4.1 and 4.9, respectively, and the optical transition at 1\,eV is essentially a charge 
transfer (CT) excitation from the occupied V 3{\em d} state at one side of the rung to the 
empty 3{\em d} state at the opposite side of the same rung.

Let us now turn to the infrared continuum detected for $E\!\perp$\,chain. The presence of 
{\em two} V states ($|L\rangle$ and $|R\rangle$) per spin, along with the broken left-right 
parity of the ground state on each rung, gives rise to the fascinating behavior of the spin flips 
described in detail in section\ \ref{chmgmod} for the 1/4 filled single two-leg ladder, and 
in a pictorial way for $\alpha^{\prime}$-NaV$_2$O$_5$, in Fig.\ \ref{nvsflip}:  
If in a small fragment of the ladder each spin resides in a $|L\rangle$ orbital, with some 
admixture of $|R\rangle$, the inclusion of the coupling between the rungs, via the 
Hamiltonian $H_{\parallel}$ (Eq.\ \ref{hpara}), has no effect when the spins are parallel, due 
to the Pauli-principle. If the spins form an $S\!=\!0$ state, the ground state gains some 
kinetic energy along the legs by mixing in states where two electrons reside on the same rung. 
In the limit  $U\!\rightarrow\!\infty$, these states have one electron in the $|L\rangle$ and 
one in the $|R\rangle$ state on the same rung. As a result, there is a net dipole 
displacement of the singlet state compared to the triplet state: 
{\em The spin-flip excitations carry a finite electric dipole moment along the rung direction.}

From the expression for the exchange coupling constant calculated in the limit 
$U\!\rightarrow\!\infty$ (Eq.\ \ref{jpara}), and using the values of $t_{\bot}$ and $\Delta$ 
calculated from the optical data and $|t_{\parallel}|\!\approx\!0.2$\,eV,\cite{harrison} we 
obtain $J_{\parallel}\!\approx\!30$ meV. It is significant that this value is comparable to those 
reported on the basis of magnetic susceptibility measurements, which range from 38 to 48 meV 
(see Ref.\ \onlinecite{weiden} and\ \onlinecite{isobe}, respectively). Moreover, for the effective 
charge of the bi-magnon excitation process (Eq.\ \ref{efcgmg}), we obtain $q_m/q_e\!=\!0.07$. 

We can also compare the experimental intensity of the spin-fluctuations relative to the CT 
excitations to the maximum limiting value calculated for this quantity (Eq.\ \ref{lsigCT/S}). 
For the parameters obtained from the conductivity spectra, the maximum relative 
intensity is $\sim$0.0014.  The experimental value is $\sim$0.0008, at T=300 K, in good agreement 
with the numerical estimate. 

We can now try to interpret the temperature dependence of the spin-fluctuation continuum 
shown in Fig.\ \ref{nvgapst}b: The oscillator strength increases upon cooling 
the sample from room temperature to $T_{\rm{c}}$, and rapidly decreases for 
$T\!<\!T_{\rm{c}}$. Within the charged-magnon model, the increase marks an increase in short
range AF correlations of the chains: In fact, the intensity of the spin-fluctuations is   
proportional to the spin-spin correlation function $g_S(T)$ defined in Eq.\ \ref{spincf}. Below 
$T_{\rm{c}}$, nn spin-singlet correlations become dominant, and $g_{S}$ is suppressed, 
along with the intensity of the spin-fluctuations. 

In conclusion, by a detailed analysis of the optical conductivity, we provided a direct evidence 
for a local breaking of symmetry in the electronic charge distribution over the rungs of the 
ladders in $\alpha^{\prime}$-NaV$_2$O$_5$, in the high temperature undistorted phase. Moreover, we showed that a direct  
two-magnon optical absorption process is responsible for the low frequency continuum observed  
perpendicularly to the V-O chains. 

\section{Discussion}
\label{nvdisc}

As we discussed in the course of this paper, the interpretation of the phase transition in 
$\alpha^{\prime}$-NaV$_2$O$_5$ is still controversial. On the one hand, the SP picture was suggested by preliminary 
magnetic susceptibility\cite{isobe} and inelastic neutron scattering measurements.\cite{fujii} 
On the other, x-ray diffraction redeterminations of the crystal structure of the compound, in 
the high temperature undistorted phase,\cite{meetsma,schnering,smolinski} showed that 
the dimerization within well defined 1D magnetic chains is not very probable in $\alpha^{\prime}$-NaV$_2$O$_5$. Moreover, 
the magnetic field dependence of $T_{\rm{c}}$, expected for a true SP system on the basis of 
Cross and Fisher theory,\cite{crossfi,cross} has not been observed.\cite{buchner,schnelle} 

Additional information came recently from NMR experiments:\cite{ohama} Above $T_{\rm{c}}$, 
only one kind of V site (magnetic) was clearly detected, whereas no signature of nonmagnetic 
V$^{5+}$ was found. However, another set of magnetic V sites which are invisible in the NMR 
spectrum could not be ruled out.\cite{ohama} On the other hand, below the phase transition, two 
inequivalent sets of V sites were detected and assigned to V$^{4+}$ and  V$^{5+}$. These results 
were then interpreted as an indication of the charge ordering nature of the phase transition, in 
$\alpha^{\prime}$-NaV$_2$O$_5$.\cite{ohama}  Subsequently, many theoretical studies have been addressing this 
possibility.\cite{seo,thalmeier,mostovoy,riera} The main point to be explained is why a 
charge ordering would automatically result in the opening of a spin gap (with the same transition 
temperature).  Among all the models, the more interesting seem to be those were a `zig-zag' charge 
ordering on the ladders is realized. In this case, the ordering would immediately result in 
a nonmagnetic ground state, due to the formation of spin singlets on nn V sites 
belonging to two different ladders\cite{seo} or, alternatively, to the modulation of 
$J_{\parallel}$ within a single ladder.\cite{mostovoy} 

In the discussion of our spectra of the optical conductivity, we showed that the infrared 
continuum, observed perpendicularly to the V-O chains direction, could be attributed to direct 
optical absorption of double spin-flip excitations. However, in order to have optical activity for 
these processes and, at the same time, to explain the energy position and the polarization 
dependence of the on-rung bonding-antibonding transition at 1\,eV, we had to assume a charge 
disproportionation between the two V sites on a rung. This assumption seems to contrast 
the results of the x-ray diffraction analysis\cite{meetsma,schnering,smolinski} and of NMR 
measurements.\cite{ohama} As far as the former is concerned, it is obvious that if the charge 
disproportionation does not have a long range character it would not be detectable in an 
x-ray diffraction experiment, but would still result in optical activity of charged bi-magnon 
excitations.  
Regarding the disagreement with the NMR results,\cite{ohama} we saw that the existence of another 
kind of V sites could not be completely excluded in the NMR spectrum.\cite{ohama} Furthermore, 
in comparing the optical to the NMR data, there could also be  a problem with the different 
time scales of the two experimental techniques (i.e., `fast' and `slow' for optics and NMR, 
respectively).  Therefore, if there are charge fluctuations in the system, as one would expect  
for a charge ordering phase transition, the experimental results could be different.

From the analysis of the optical conductivity, we cannot really say which is the more appropriate 
picture for the phase transition in $\alpha^{\prime}$-NaV$_2$O$_5$. We saw that, indeed, the 
electronic and/or magnetic degrees of freedom are playing a leading role. This is shown, e.g., by 
the strong temperature dependence of the infrared spin fluctuation spectrum which, in turn, 
determines the anomalous behavior of the microwave dielectric constant.\cite{sekine,smirnov5} 
However, the reduction of the oscillator strength of the infrared continuum for 
$T\!<\!T_{\rm{c}}$, which we discussed as a consequence of 
nn spin-singlet correlations becoming dominant below $T_{\rm{c}}$, could also result from a 
zig-zag charge ordering on the ladders. In this case, as for the perfectly symmetrical 
1/4\,-filled ladder, the optical activity of double spin flips would simply vanish.

In our opinion, for a better understanding of the phase transition, the pretransitional 
fluctuations observed with optical spectroscopy, ultrasonic wave propagation,\cite{fertey} 
and x-ray diffuse scattering,\cite{ravy} have to be taken into account. They probably reflect the 
high degree of disorder present in the system above the phase transition, which is also shown 
by the local and `disordered' charge disproportionation necessary to interpret the optical 
conductivity. In this context, a charge ordering transition, where more degrees of freedom than 
just those of the spins are involved, seems plausible. It would also explain the experimentally 
observed entropy reduction,  which appeared to be larger than what expected for a 1D AF 
HB chain.\cite{buchner} 

Also the degeneracy of the zone boundary folded modes, activated by the phase 
transition, could be an important piece of information. However, at this stage of the investigation 
of the electronic and magnetic properties of $\alpha^{\prime}$-NaV$_2$O$_5$, very important insights could come from the 
x-ray diffraction determination of the low temperature structure. Probably, only at that point 
it will be possible to fit all the different results together and come up with the 
final interpretation of the phase transition.

\section{Conclusions}

In this paper we have been discussing the high and low temperature optical properties of 
$\alpha^{\prime}$-NaV$_2$O$_5$, in the energy range 4\,meV-4\,eV. We studied the symmetry of the 
material, in the high temperature undistorted phase, comparing the different  crystal structures 
proposed on the basis of x-ray diffraction\cite{carpy,meetsma,schnering,smolinski} to our infrared 
phonon spectra. We found that the system is better described by the centrosymmetric space group 
{\em Pmmn}, in agreement with the recent structure 
redetermination.\cite{meetsma,schnering,smolinski} 

On the other hand, by a detailed analysis of the electronic excitations detected in the 
optical conductivity, we provided direct evidence for a charged disproportionated electronic 
ground-state, at least on a locale scale: A consistent interpretation of both structural and 
optical conductivity data requires an asymmetrical charge distribution on each rung, 
without any long range order. We showed that, 
because of the locally broken symmetry, spin-flip excitations carry a finite electric 
dipole moment,  which is responsible for the detection  of charged bi-magnons in the optical 
spectrum for $E\!\perp$\,chain, i.e., direct two-magnon optical absorption processes.

By analyzing the optically  allowed phonons at various temperatures below and above the phase 
transition, we concluded that a second-order change to a larger unit cell takes place below 34 K, 
with a fluctuation regime  extending over a very broad temperature range.

The charged-magnon model we have been discussing in this paper has been developed explicitly to 
interpret the peculiar optical conductivity spectra of 
$\alpha^{\prime}$-NaV$_2$O$_5$. However, our opinion is that the importance of this  
model goes beyond the understanding of the excitation spectrum of just this particular system. 
We expect this model (obviously in a more general and rigorous form), to be relevant to many of 
the strongly correlated electron systems, where the interplay of spin and charge plays 
a crucial role in determining the low energy electrodynamics.

At this point we would like to speculate that a specific case where charged magnon excitations 
could be observed is the stripe phase of copper oxides 
superconductors\cite{tranquada} and of other strongly correlated oxides.\cite{chen} In fact, 
even though the magnetism is quasi 2D in these systems, in the stripe phase a symmetry-breaking 
quasi 1D charge 
ordering takes place, which could result in optical activity for the charged magnons. For 
instance, in the optical conductivity spectra of  
La$_{1.67}$Sr$_{0.33}$NiO$_4$ 
below the charge ordering transition,\cite{katsufuji} it is possible to observe some residual 
conductivity inside the energy gap, whose nature has not been completely understood so far. 
Without further experimental and theoretical investigations it is of course not possible to make 
a strong claim. However, the reported 
features are suggestive of charged magnon excitations. In fact, for the value of $J$ obtained from 
the two-magnon Raman scattering on this material,\cite{yamamoto} direct two-magnon optical 
absorption is expected in the same energy range (i.e., from zero to $\sim$0.2 eV) where 
the residual structure has been experimentally observed.\cite{katsufuji} 

Pushing the speculation to its limit, one could imagine that even the local breaking of symmetry 
introduced by a single impurity in an AF could result in optical activity for pure spin-flip 
excitations. One particular system where this situation is probably 
realized is Zn-substituted YBa$_2$Cu$_3$O$_6$, where  Zn is introduced in the CuO$_2$ planes.   
In the cuprate parent compounds direct bi-magnon optical absorption 
is not allowed due to the inversion symmetry; only phonon-assisted magnetic 
excitations are optically active, whenever the involved phonon is lowering the symmetry of the 
system. In fact, phonon assisted bi-magnon excitations in the mid-infrared region were detected 
on both YBa$_2$Cu$_3$O$_6$ and YBa$_2$Cu$_{2.85}$Zn$_{0.15}$O$_6$.\cite{markus1} But on the  
Zn substituted system it is possible to observe, in the optical conductivity 
spectra, various features with frequencies coincident with those of pure double spin-flip 
excitations and, therefore, most probably ascribable to direct charged bi-magnon absorption 
processes.\cite{markus1}

\section{Acknowledgements}

We gratefully acknowledge M. Mostovoy, D.I. Khomskii, T.T.M. Palstra, and G.A. Sawatzky for 
stimulating discussions. We thank D. Smirnov, J. Leotin, M. Gr\"uninger, P.H.M. van Loosdrecht, 
and M.J. Rice for many useful comments, and C. Bos, A. Meetsma, and J.L. de Boer for assistance. 
One of us (A.D.) is greatful to M. Picchietto and B. Top$\acute{\rm{\i}}$ 
for their unlimited co-operation.
This investigation was supported by the Netherlands Foundation for Fundamental Research on 
Matter (FOM) with financial aid from the Nederlandse Organisatie voor Wetenschappelijk 
Onderzoek (NWO).


\begin{references}
\bibitem[*]{byline}Present address: Center for Materials Research, McCullough Building, 
          Stanford University, 476 Lomita Mall, Stanford, CA 94305-4045. 
           E-mail: damascel@stanford.edu
\bibitem{isobe}M. Isobe, and Y. Ueda, J.\ Phys.\ Soc.\ Jpn.\ {\bf65}, 1178 (1996).
\bibitem{fujii}Y. Fujii, H. Nakao, T. Yosihama, M. Nishi, K. Nakajima, K. Kakurai,
                  M. Isobe, Y. Ueda, and H. Sawa,  J.\ Phys.\ Soc.\ Jpn.\ {\bf66}, 326 (1997).
\bibitem{carpy}P.A. Carpy, and J. Galy, Acta\ Cryst.\ B\ {\bf31}, 1481 (1975).
\bibitem{nishi}M. Nishi, O. Fujita, and J. Akimitsu, Phys.\ Rev.\ B\ {\bf50}, 6508 (1994).
\bibitem{uhrig}G.S. Uhrig, Phys.\ Rev.\ Lett.\ {\bf79}, 163 (1997).
\bibitem{buchner}B. B\"uchner, private communication.
\bibitem{schnelle}W. Schnelle, Yu. Grin, and R.K. Kremer, Phys.\ Rev.\ B\ {\bf59}, 73 (1999).
\bibitem{crossfi}M.C. Cross, and D.S. Fisher, Phys.\ Rev.\ B\ {\bf19}, 402 (1979).
\bibitem{cross} M.C. Cross, Phys.\ Rev.\ B\ {\bf20}, 4606 (1979).
\bibitem{meetsma}A. Meetsma, J.L. de Boer, A. Damascelli, J. Jegoudez, A. Revcolevschi,
                    and T.T.M. Palstra, Acta\ Cryst.\ C\ {\bf54}, 1558 (1998).
\bibitem{schnering}H.G. von Schnering, Yu. Grin, M. Kaupp, M. Somer, R.K. Kremer, and O. Jepsen, 
                        Z.\ Kristallogr.\ {\bf213}, 246 (1998).
\bibitem{smolinski}H. Smolinski, C. Gros, W. Weber, U. Peuchert, G. Roth, M. Weiden, 
                       and C. Geibel, Phys.\ Rev.\ Lett.\ {\bf80}, 5164 (1998).
\bibitem{anvprl}A. Damascelli, D. van der Marel, M. Gr\"uninger, C. Presura,  T.T.M. Palstra,
                    J. Jegoudez, and A. Revcolevschi, Phys.\ Rev.\ Lett.\ {\bf81}, 918 (1998). 
\bibitem{anvphysica}A. Damascelli, D. van der Marel, J. Jegoudez, G. Dhalenne, 
                       and A. Revcolevschi, Physica\ B\ {\bf259-261}, 978 (1999).
\bibitem{ohama}T. Ohama, H. Yasuoka, M. Isobe, and Y. Ueda, Phys.\ Rev.\ B {\bf59}, 3299 (1999).  
\bibitem{seo}H. Seo, and H. Fukuyama, J.\ Phys.\ Soc.\ Jpn.\ {\bf67}, 2602 (1998).
\bibitem{thalmeier}P. Thalmeier, and P. Fulde, Europhys.\ Lett.\ {\bf44}, 242 (1998).
\bibitem{mostovoy}M. Mostovoy, and D. Khomskii, cond-mat/9806215 (18 June 1998).
\bibitem{riera}J. Riera, and D. Poilblanc, Phys.\ Rev.\ B\ {\bf4}, 2667 (1999). 
\bibitem{bonner}J.C. Bonner, and M.E. Fisher, Phys.\ Rev.\ A\ {\bf135}, 610 (1964).
\bibitem{mila}F. Mila, P. Millet, and J. Bonvoisin, Phys.\ Rev.\ B\ {\bf54}, 11\,925 (1996).
\bibitem{isobe1}M. Isobe, and Y. Ueda,  Physica\ B\ {\bf262}, 180 (1997).
\bibitem{hemberger}J. Hemberger, M. Lohmann, M. Nicklas, A. Loidl, M. Klemm, G. Obermeier, and 
                     S. Horn, Europhys.\ Lett.\ {\bf42}, 661 (1998).
\bibitem{rousseau}D.L. Rousseau, R.P. Bauman, and S.P.S. Porto, J.\ Raman\ Spectrosc.\ 
                      {\bf10}, 253 (1981).
\bibitem{perkins}J.D. Perkins, J.M. Graybeal, M.A. Kastner, R.J. Birgeneau, J.P. Falck, 
        and M. Greven, Phys.\ Rev.\ Lett.\ {\bf71}, 1621 (1993).
\bibitem{lorenzana1}J. Lorenzana, and G.A. Sawatzky, Phys.\ Rev.\ Lett.\ {\bf74}, 1867 (1995).
\bibitem{lorenzana2}J. Lorenzana, and G.A. Sawatzky, Phys.\ Rev.\ B\ {\bf52}, 9576 (1995). 
\bibitem{suzuura}H. Suzuura, H. Yasuhara, A. Furusaki, N. Nagaosa, and Y. Tokura, 
         Phys.\ Rev.\ Lett.\ {\bf76}, 2579 (1996).
\bibitem{lorenzana3}J. Lorenzana, and R. Eder, Phys.\ Rev.\ B\ {\bf55}, R3358 (1997).
\bibitem{kastner}For a review, see M.A. Kastner, R.J. Birgeneau, G. Shirane, and Y. Endoh,         
         Rev.\ Mod.\ Phys.\ {\bf70}, 897 (1998), and references therein.
\bibitem{ziman}J.M. Ziman, {\em Principles of the Theory of Solids} 
                (Cambridge University Press, Cambridge, 1972).
\bibitem{mjr79}M.J. Rice, Solid\ State\ Comm.\ {\bf31}, 93 (1979).
\bibitem{faddeev}L.D.Faddeev, and L.A. Takhtajan, Phys.\ Lett.\ A\  {\bf85}, 744 (1981).
\bibitem{ueda}M. Isobe, C. Kagami, and Y. Ueda, J.\ Crystal\ Growth\  {\bf181}, 314 (1997).
\bibitem{golubchik}A. Golubchik, M. Isobe, A.N. Ivlev, B.N. Mavrin, M.N. Popova, A.B. Sushkov, 
                       Y. Ueda, and A.N. Vasil'ev, J.\ Phys.\ Soc.\ Jpn.\ {\bf66}, 4042 (1997).
\bibitem{smirnov}D. Smirnov, J. Leotin, P. Millet, J. Jegoudez, and A. Revcolevschi, 
                     Physica\ B\ {\bf259-261}, 992 (1999).
\bibitem{popova1}M.N. Popova, A.B. Sushkov, A.N. Vasil'ev, M. Isobe, and Y. Ueda, 
                JETP\ Lett.\ {\bf65}, 743 (1997).
\bibitem{smirnovprb}D. Smirnov, P. Millet, J. Leotin, D. Poilblanc, J. Riera, D. Augier, and 
                P. Hansen, Phys.\ Rev.\ B\ {\bf57}, R11\,035 (1998).
\bibitem{popovic}Z.V. Popovi$\rm{\acute{c}}$, M.J. Konstantinovi$\rm{\acute{c}}$, R. 
                Gaji$\rm{\acute{c}}$, V. Popov, Y.S. Raptis, A.N. Vasil'ev, M. Isobe, and Y Ueda, 
                 J.\ Phys.\ Condens.\ Matter\ {\bf10}, 513 (1998).
\bibitem{fischer}M. Fischer, P. Lemmens, G. G\"untherodt, M. Weiden, R. Hauptmann, C. Geibel, 
                        and F. Steglich, Physica\ B\ {\bf244}, 76 (1998). 
\bibitem{lemmens}P. Lemmens, M. Fischer, G. Els, G. G\"untherodt, A.S. Mishchenko, M. Weiden, 
                R. Hauptmann, C. Geibel, and F. Steglich, Phys.\ Rev.\ B\ {\bf58}, 14\,159 (1998). 
\bibitem{kuroe}H. Kuroe, H. Seto, J. Sasaki, T.Sekine, M. Isobe, and Y. Ueda, 
               J.\ Phys.\ Soc.\ Jpn.\ {\bf67}, 2881 (1998).
\bibitem{popova2}M.N. Popova, A.B. Sushkov, S.A. Golubchik, B.N. Mavrin, V.N. Denisov, B.Z. 
                Malkin, A.I. Iskhakova, M. Isobe, and Y. Ueda, cond-mat/9807369 (28 July 1998).
\bibitem{fano}U. Fano, Phys.\ Rev.\ {\bf124}, 1866 (1961).
\bibitem{ludecke}J. L\"udecke, A. Jobst, S. van Smaalen, E. Morr$\acute{\rm{e}}$, C. Geibel, 
                and H.-G. Krane, Phys.\ Rev.\ Lett.\ {\bf82}, 3633 (1999). 
\bibitem{sp1}A. Damascelli, D. van der Marel, F. Parmigiani, G. Dhalenne, and A. Revcolevschi, 
                Phys.\ Rev.\ B\ {\bf56}, R11\,373 (1997).
\bibitem{sp2}A. Damascelli, D. van der Marel, F. Parmigiani, G. Dhalenne, and A. Revcolevschi, 
                 Physica\ B\  {\bf244}, 114 (1998).
\bibitem{sp3}A. Damascelli, D. van der Marel, G. Dhalenne, and A. Revcolevschi, 
                submitted to Phys.\ Rev.\ B (1999).
\bibitem{fesi}A. Damascelli, K. Schulte, D. van der Marel, and A.A. Menowsky, 
                  Phys.\ Rev.\ B\ {\bf55}, R4863 (1997).
\bibitem{fertey}P. Fertey, M. Poirier, M. Castonguay, J. Jegoudez, and A. Revcolevschi, 
                    Phys.\ Rev.\ B\ {\bf57}, $13\,698$ (1998).
\bibitem{ravy}S. Ravy, J. Jegoudez, and A. Revcolevschi, Phys.\ Rev.\ B {\bf2}, R681(1999). 
\bibitem{bassi}M. Bassi, P. Camagni, R. Rolli, G. Samoggia, F. Parmigiani, G. Dhalenne, 
                   and A. Revcolevschi , Phys.\ Rev.\ B.\ {\bf54}, R$11\,030$ (1996).
\bibitem{horsch}P. Horsch, and F. Mack, Eur.\ Phys.\ J.\ B {\bf5}, 367 (1998).
\bibitem{nishimoto}S. Nishimoto, and Y. Ohta, J.\ Phys.\ Soc.\ Jpn.\ {\bf67}, 2996 (1998). 
\bibitem{weiden}M. Weiden, R. Hauptmann, C. Geibel, F. Steglich, M. Fischer, P. Lemmens, 
                    and G. G\"untherodt, Z.\ Phys.\ B\ {\bf103}, 1 (1997).
\bibitem{sekine}Y. Sekine, N. Takeshita, N. M$\hat{\rm{o}}$ri, M.Isobe, and Y.Ueda, 
                  preprint (1998).
\bibitem{smirnov5}A.I.Smirnov, M.N.Popova, A.B.Sushkov, S.A.Golubchik, D.I.Khomskii, 
                    M.V.Mostovoy, A.N.Vasil'ev, M.Isobe, and Y.Ueda, 
                     cond-mat/9808165 (27 October 1998).
\bibitem{nishimoto1}S. Nishimoto, and Y. Ohta, J.\ Phys.\ Soc.\ Jpn.\ {\bf67}, 3679 (1998).
\bibitem{harrison}W.A. Harrison, {\em Electronic Structure and the Properties of Solids} 
(Dover Publications, New York, 1989).
\bibitem{tranquada}J.M. Tranquada, B.J. Sternlieb, J.D. Axe, Y. Nakamura, and S. Uchida, 
        Nature {\bf375}, 561 (1995).
\bibitem{chen}C.H. Chen, S-W. Cheong, and A.S. Cooper, Phys.\ Rev.\ Lett.\ {\bf71}, 2461 (1993).
\bibitem{katsufuji}T. Katsufuji, T. Tanabe, T. Ishikawa, Y. Fukuda, T. Arima, and Y. Tokura,
        Phys.\ Rev.\ B\ {\bf54}, R$14\,230$ (1996).         
\bibitem{yamamoto}K. Yamamoto, T. Katsufuji, T. Tanabe, and Y. Tokura , 
        Phys.\ Rev.\ Lett.\ {\bf80}, 1493 (1998).
\bibitem{markus1}M. Gr\"uninger, D. van der Marel, A. Damascelli, A. Zibold, H.P. Geserich, 
        A. Erb, M. Kl\"aser, Th. Wolf, T. Nunner, and T. Kopp, Physica C, in press (1999).
\end{references}
\end{document}